\documentclass[aps,prb,reprint,superscriptaddress,floatfix]{revtex4-2}
\usepackage{amsmath,amssymb,amsfonts}
\usepackage{slashed}
\usepackage{graphicx}
\usepackage{bm}
\usepackage{hyperref}
\usepackage{booktabs}

\newcommand{\kxy}{\kappa_{xy}/T}
\newcommand{\Kem}{K_{\rm em}}

\newcommand{\Dpg}{\Delta_{\rm pg}}
\newcommand{\Dsc}{\Delta_{\rm sc}}

\newcommand{\Tstar}{T^{*}}

\newcommand{\vth}{\vartheta}
\newcommand{\Leff}{L_{\rm eff}}
\newcommand{\R}{\mathbb{R}}

\newcommand{\dd}{\mathrm{d}}
\newcommand{\tr}{\mathrm{tr}}
\newcommand{\sgn}{\mathrm{sgn}}
\newcommand{\CBdG}{C_{\rm BdG}}

\begin{document}

\title{Parity anomaly governs the thermal Hall response of chiral
superconductivity in rhombohedral graphene above and below $T_c$}

\author{Kumar Ghosh}
\email{jb.ghosh@outlook.com}
\affiliation{E.ON Digital Technology, Laatzener Str.\ 1,
             30539 Hannover, Germany}

\begin{abstract}
Chiral superconductivity has recently been confirmed in rhombohedral
tetra- and pentalayer graphene near the BCS--BEC crossover, but no
theory tells the experimentalist what thermal Hall response to
expect, or why any signal should persist above the phase-coherence
temperature $T_c$.  We show that the parity anomaly of
$(2{+}1)$-dimensional field theory fixes the answer exactly, at all
temperatures, with no free parameters:
$\kappa_{xy}/T = (\pi^2 k_B^2/6h)\,\CBdG\,\tanh[\Delta(T)/(2k_BT)]$,
where $\CBdG$ is the BdG Chern number and $\Delta(T)$ is the
fermionic excitation gap.  The BCS--BEC two-gap relation
$\Delta^2 = \Dsc^2 + \Dpg^2$ makes the same formula govern both the
condensate and pseudogap regimes, so the signal onsets at the
pair-formation temperature $\Tstar$ rather than at $T_c$.
Coleman--Hill non-renormalization and the $c_1 = 0$ theorem protect
this result against interactions and finite-size artefacts.  Three
independent numerical validations confirm the topological input at
machine precision (FHS Chern numbers, Wilson-loop $c_1 = 0$ test)
and at the many-body level (DMRG on 28 converged ground states,
including the real-space $p+ip$ signature $\arg\mathcal{A}_y - \arg\mathcal{A}_x = -\pi/2$ recovered to $10^{-14}$).  The theory delivers an immediate falsifiable test that requires no new experimental apparatus: the sign of $\kappa_{xy}/T$ below $T_c$ must equal the sign of the anomalous Hall resistance $R_{xy}$ already measured above $T_c$, and four further predictions accessible by dilution-refrigerator nano-calorimetry on existing devices.
\end{abstract}

\maketitle

\section{Introduction}
\label{sec:intro}

The recent discovery of chiral superconductivity in rhombohedral
tetralayer and pentalayer graphene by Han et al.~\cite{Han2025} is a
landmark for topological superconductivity in a pure carbon
platform.  Two superconducting states (SC1 and SC2) appear with
$T_c$ up to $300$~mK, with spontaneous time-reversal breaking
established by hysteretic $R_{xx}$, in-plane-field robustness, and
zero-field anomalous Hall signals in the normal state.  The gap
ratio $2\Delta/k_B T_c \sim 10\text{--}30$ places the system near
the BCS--BEC crossover~\cite{Han2025,Chen2024RMP}, and the
anomalous Hall angle $\tan\theta_H = R_{xy}/R_{xx} \approx 0.1$
above $T_c$~\cite{Han2025} quantifies the Berry curvature of the
valley-polarized normal state that selects the pairing chirality.

Two questions immediately follow, and no existing theory answers
either.  \emph{First}, what does the thermal Hall response
$\kappa_{xy}/T$ look like as a function of temperature?  The
standard Kubo calculation~\cite{LeNir2026} delivers only the
$T \to 0$ quantized plateau, but any measurement at finite
temperature (i.e., every measurement) needs the full profile.
\emph{Second}, does anything survive above $T_c$?  In the BCS--BEC
crossover regime the pseudogap $\Dpg(T)$ persists to a
pair-formation scale $\Tstar \gg T_c$, so a fermionic gap remains
after phase coherence is lost.  Whether the thermal Hall response
tracks that gap or vanishes with the condensate is a physical
question with no existing theoretical answer for any material,
let alone a graphene device where the experiment is imminent.

\begin{figure*}[t]
  \centering
  \includegraphics[width=0.9\textwidth]{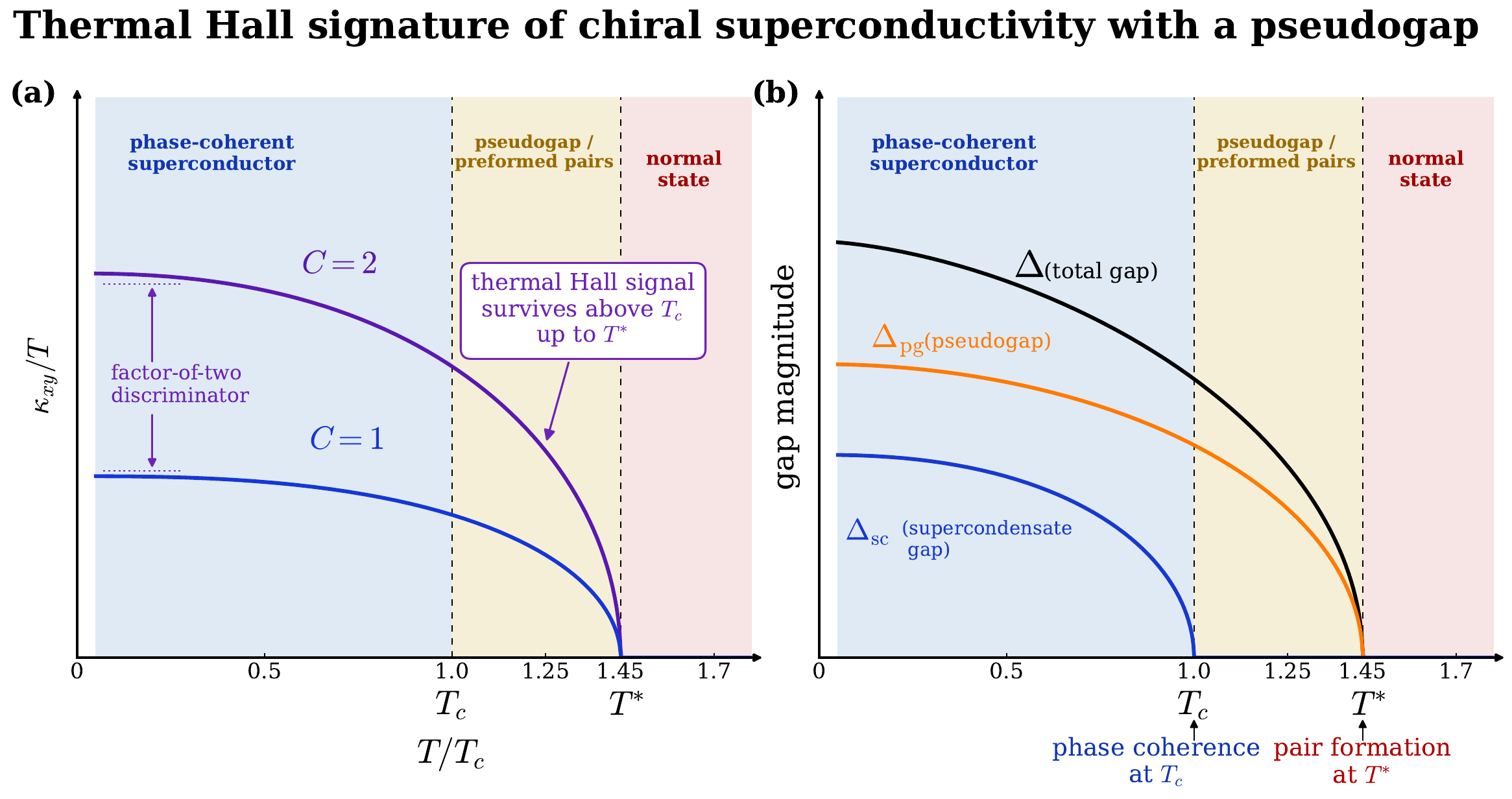}
  \caption{Central prediction.  (a)~Temperature dependence of the
    anomaly-fixed thermal Hall conductance
    $\kappa_{xy}/T = (\pi^2 k_B^2/6h)\,\CBdG\,\tanh[\Delta(T)/2k_BT]$
    for $\CBdG = 1$ and $\CBdG = 2$: the two states are separated by
    a factor of two at low temperature, and the signal is continuous
    through $T_c$ and terminates at $\Tstar$, not at $T_c$.
    (b)~BCS--BEC two-gap structure: the total fermionic gap
    $\Delta^2 = \Dsc^2 + \Dpg^2$ that enters the anomaly formula is
    controlled by the condensate $\Dsc$ below $T_c$ and by the
    pseudogap $\Dpg$ up to $\Tstar$.}
  \label{fig:fig1_schematic}
\end{figure*}

We resolve both questions with a single closed-form result.  The
holonomy-resummed parity-odd
kernel~\cite{GhoshKlinkhamer2017,Ghosh2026,Ghosh2026BEC}, combined
with the gravitational Chern--Simons
relation~\cite{Ryu2012,Volovik2003}, fixes the thermal Hall
conductance at every temperature with no free parameters:
\begin{equation}
  \boxed{
  \frac{\kappa_{xy}}{T}
  = \frac{\pi^2 k_B^2}{6h}\,\CBdG\,
    \tanh\!\left[\frac{\Delta(T)}{2k_BT}\right],
  }
\label{eq:kxy-main}
\end{equation}
where $\CBdG$ is the BdG Chern number and $\Delta(T)$ is the
fermionic excitation gap.  The formula has three features that no
prior treatment provides.  (i) It is exact at finite $T$, not
merely at $T \to 0$.  (ii) The same expression governs the
condensate and pseudogap regimes: below $T_c$, $\Delta(T)$ is the
BdG gap; above $T_c$, the BCS--BEC two-gap
relation~\cite{Chen2024RMP}
$\Delta^2 = \Dsc^2 + \Dpg^2$ makes $\Delta(T > T_c) = \Dpg(T)$ the
parity-anomaly mass, so the signal onsets at $\Tstar$ rather than
at $T_c$ (Fig.~\ref{fig:fig1_schematic}).  (iii)~Coleman--Hill
non-renormalization~\cite{ColemanHill1985} shields the map from
gap to Chern--Simons level from higher-loop corrections, and the
$c_1 = 0$ theorem~\cite{Ghosh2026,Ghosh2026BEC} guarantees purely
exponential (not power-law) finite-size corrections.  Together
these give a formula that survives both interactions and finite
geometry.

Three independent numerical validations confirm the topological
input.  Machine-precision Fukui--Hatsugai--Suzuki Chern-number
calculations on the square-lattice $p+ip$ anchor and on the
RHG-effective Lifshitz model recover $\CBdG = 1$ and $\CBdG = 2$ at
integer error $\lesssim 10^{-14}$.  Wilson-loop flux threading on
finite cylinders confirms the $c_1 = 0$ exponential envelope
directly.  DMRG on 28 converged ground states, ranging
$L_y \in \{4,6,8,10\}$ and $\Dpg \in \{0.2, \ldots, 0.8\}$,
delivers six independent many-body signatures of the $\CBdG = 1$
chiral $p+ip$ phase; the sharpest is the real-space pairing-phase
signature
$\arg\langle c_i c_{i+\hat y}\rangle
 - \arg\langle c_i c_{i+\hat x}\rangle = -\pi/2$
recovered to $2.2 \times 10^{-14}$ at $L_y = 4$.
The formula translates to five falsifiable experimental
predictions, of which the sharpest requires no new apparatus: the
sign of $\kappa_{xy}/T$ below $T_c$ must equal the sign of the
anomalous Hall $R_{xy}$ already measured above $T_c$, because both
are locked to the same Berry-curvature sign of the parent
quarter-metal band~\cite{li2025berrytrashcan}.

\emph{Organization.}  Section~\ref{sec:bdg-model} specifies the BdG
Hamiltonian and computes $\CBdG$ for each pairing scenario.
Section~\ref{sec:anomaly} derives Eq.~\eqref{eq:kxy-main} from the
holonomy-resummed kernel.
Section~\ref{sec:pseudogap} extends the formula through $T_c$ via
the two-gap relation.
Section~\ref{sec:numerical} presents the three-tier numerical
validation.
Section~\ref{sec:predictions} lists the five falsifiable
predictions, ordered by experimental immediacy.

\section{BdG Model and Chern Number for Rhombohedral Graphene}
\label{sec:bdg-model}

\subsection{Band dispersion and pairing Hamiltonian}
\label{subsec:dispersion}

The superconducting state SC1 develops within a spin- and
valley-polarized quarter-metal phase, a fully isospin-polarized
normal state with a single Fermi pocket near one $K$
valley~\cite{Ghazaryan2023,Geier2025}.  The band dispersion is
strongly tunable by the displacement field $D$ and can undergo a
Lifshitz transition from a simply-connected to an annular Fermi
sea; the FFT Landau-fan spectrum of the putative parent state of
SC2 shows a diagonal feature above filling $f_v = 1$~\cite{Han2025},
the fingerprint of a quarter-metal with annular Fermi surface.
Several proposals identify the pairing symmetry as chiral
$p_x + i\tau p_y$ ($\tau = \pm 1$ for $K/K'$), driven by
overscreened Coulomb repulsion~\cite{Geier2025} or related
mechanisms, with the chirality selected by the Berry curvature of
the valley-polarized band~\cite{Geier2025,LeNir2026,Patri2025}.
The BdG Chern number $\CBdG$ of the superconducting state is the
central topological invariant of the problem, and the thermal Hall
conductance $\kappa_{xy}/T$ is its most direct
probe~\cite{Read2000,Kane1997}.  Le Nir, Mitra, and
Kim~\cite{LeNir2026} recently computed $\CBdG$ and the $T \to 0$
thermal Hall conductivity for chiral superconductors from Chern
parent bands motivated by rhombohedral graphene, establishing that
$\CBdG$ equals the sum of momentum-space vortex windings enclosed
within the occupied Fermi sea (Eq.~\eqref{eq:CBdG-vortex} below).

The low-energy dispersion near the $K$ valley in the quarter-metal
state is captured by the minimal model~\cite{Ghazaryan2023,Geier2025}:
\begin{equation}
  \varepsilon_k = D\sqrt{1 + (k/k_0)^{2N}} + \frac{\hbar^2 k^2}{2m},
\label{eq:dispersion}
\end{equation}
where $D$ is the displacement-field-induced gap, $k_0$ and $m$ are
$D$-dependent parameters, and $N$ is the number of layers.  For
$N = 4$ (tetralayer) and $N = 5$ (pentalayer), the dispersion has a
Mexican-hat shape at large $D$, producing an annular Fermi surface via
a Lifshitz transition~\cite{Geier2025}.

The BdG Hamiltonian for a chiral $p + ip$ superconductor takes the
standard Nambu form:
\begin{equation}
  H_{\rm BdG}(\bm{k})
  = \begin{pmatrix}
      \xi_{\bm{k}} & \Delta_{\bm{k}} \\
      \Delta_{\bm{k}}^* & -\xi_{-\bm{k}}
    \end{pmatrix},
\label{eq:BdG-ham}
\end{equation}
with $\xi_{\bm{k}} = \varepsilon_{\bm{k}} - \mu$ and chiral gap
function $\Delta_{\bm{k}} = \Delta_0 \, (k_x + i\tau k_y)/k_F$ for
intravalley spin-triplet pairing in the quarter-metal state
($\tau = \pm 1$ labels the valley).

For numerical validation we use the square-lattice spinless $p+ip$
model as a universal topological representative:
\begin{equation}
  H_{\rm BdG}(\bm{k}) = d_x\tau_x + d_y\tau_y + d_z\tau_z,
\label{eq:pwave-bdg}
\end{equation}
with $d_x = \Delta_0\sin k_x$, $d_y = \Delta_0\sin k_y$, and
$d_z = 2t(2 - \cos k_x - \cos k_y) - \mu$.  For $0 < \mu < 4t$ the
system is in the weak-pairing topological phase with $\CBdG = 1$.

For band-projected superconductivity on a Chern
band~\cite{LeNir2026,Geier2025}, the gap function acquires
momentum-space vortices seeded by the parent-band Berry curvature,
and the BdG Chern number reduces to the total occupied vortex
charge~\cite{LeNir2026}:
\begin{equation}
  \CBdG = \mathcal{Q}
        = \sum_{i:\,\bm{k}_i \in D_{\rm occ}} \ell_i,
\label{eq:CBdG-vortex}
\end{equation}
where the sum runs over vortices of winding $\ell_i$ at positions
$\bm{k}_i$ enclosed within the occupied region of the Fermi sea.

\subsection{Chern number for each pairing scenario}
\label{subsec:chern-scenarios}

The Chern number depends on the pairing symmetry and the Fermi
surface topology.

\paragraph{Scenario A: Intravalley spin-triplet $p + ip$.}
The quarter-metal state has a single spin-valley-polarized pocket
near $K$.  For the simply-connected Fermi surface ($D < D_c$), the
gap has a single $\ell = 1$ vortex at $\bm{k} = 0$ within the Fermi
disk, giving
\begin{equation}
  \CBdG = 1.
\label{eq:C-pip}
\end{equation}
This is the Read--Green
result~\cite{Read2000}: the $p + ip$ state is a topological
superconductor with a single chiral Majorana edge mode.

The condition $\CBdG = 1$ specifically requires the Fermi surface to
lie \emph{inside} the Berry ring of fire (BRF) at radius
\begin{equation}
  k_\Omega = \left(\frac{2D^2}{v_N^2}\,
    \frac{N-1}{N+2}\right)^{1/2N},
\label{eq:BRF-radius}
\end{equation}
where the Berry curvature of R$N$G is strongly peaked and encloses
flux $-N\pi$~\cite{Patri2025}.  For $k_F > k_\Omega$, momentum-space
vortices nucleated on the BRF shift $\CBdG$ to higher odd values
$\pm(2n+1)$; the SC1 density $n_e \approx 10^{12}$~cm$^{-2}$ in Han
et al.~\cite{Han2025} falls in the $k_F < k_\Omega$ regime, confirming
$\CBdG = +1$.  The sign $\CBdG = +1$ (rather than $-1$) is fixed by
the Berry curvature of the parent quarter-metal band: the exact
many-body ground state of R$n$G under attractive
interactions~\cite{li2025berrytrashcan,LeNir2026} selects $p_x + ip_y$
because the SC chirality is ferromagnetically locked to $\sgn(\beta)$,
where $\beta > 0$ for the $K$-valley quarter-metal pocket.

For the annular Fermi surface ($D > D_c$), the central vortex at
$\bm{k} = 0$ may lie outside the occupied annulus.  If the
parent-band Berry curvature nucleates additional momentum-space
vortices within the annulus~\cite{LeNir2026,Patri2025}, $\CBdG$ can take
even values, potentially including $\CBdG = 0$ (trivial
state)~\cite{Geier2025}.  The topological phase transition at the
Lifshitz point $D = D_c$ is marked by a gap closing at
$\bm{k} = 0$~\cite{Geier2025}.

\paragraph{Scenario B: Intervalley singlet $d + id$.}
If pairing occurs between the two valleys with $d_{x^2-y^2} + id_{xy}$
symmetry, the two Dirac points at $K$ and $K'$ each contribute a
Chern number of opposite sign, giving $\CBdG = \pm 2$.

\paragraph{Scenario C: Higher Chern numbers from
$(1{+}n)$ twisted graphene.}
In $(1{+}n)$ twisted graphene structures where the superconductor
inherits a parent Chern band with $C_P = n$, the total vorticity of
the gap is $V_\Delta = 2C_P$~\cite{LeNir2026}, and $\CBdG$ can reach
$\CBdG = 2n - 1$ for a simply-connected Fermi surface.  This gives
a multi-quantum thermal Hall prediction.

The Chern-number computation is validated numerically in
Sec.~\ref{sec:numerical} using the Fukui--Hatsugai--Suzuki
discretization~\cite{Fukui2005} and Wilson-loop flux threading.

\section{Exact Finite-Temperature Chern--Simons Level and Thermal Hall}
\label{sec:anomaly}

\subsection{Holonomy-resummed parity-odd kernel}
\label{subsec:kernel}

The thermal Hall conductance of a chiral topological phase is
determined by the gravitational Chern--Simons coefficient, which at
finite temperature is controlled by the parity-odd determinant of the
massive Dirac fermions constituting the BdG quasiparticles.  We derive
the exact finite-temperature result using the holonomy-resummed kernel
of Refs.~\cite{GhoshKlinkhamer2017,Ghosh2026, Ghosh2026BEC}.

Consider a single two-component Dirac fermion of mass $m_v$ on the
Euclidean three-manifold
$\R_\tau \times \R_x \times S^1_L$ with compact coordinate
$y \equiv y + L$.  The boundary condition is
$\psi_v(y+L) = e^{i\alpha_v}\psi_v(y)$, so the total holonomy phase
is $\vth_v = \alpha_v + q_v\oint_{S^1} a_y\,\dd y$.  The
parity-odd part of the vacuum-polarization tensor takes the
transverse form~\cite{Ghosh2026}:
\begin{equation}
  \Pi_{v,\text{odd}}^{\mu\nu,\text{IR}}(p)
  = \frac{iq_v^2}{2\pi}\,
    \mathcal{K}_v^{\text{IR}}(p;L,\vth_v)\,
    \epsilon^{\mu\nu\rho}p_\rho,
\label{eq:Pi-odd}
\end{equation}
where the full mass-dependent kernel, derived by resumming all
Kaluza--Klein winding modes, is~\cite{Ghosh2026}:
\begin{equation}
  \boxed{
  \begin{aligned}
  \mathcal{K}_v^{\text{IR}}(p;L,\vth_v)
  &= \frac{\chi_v m_v}{2}\int_0^1 \frac{\dd u}{\Delta_{v,u}}\,\\
  &\quad \times \frac{\sinh(\Leff\Delta_{v,u})}
         {\cosh(\Leff\Delta_{v,u}) - \cos(\vth_v + 2\pi r u)},
  \end{aligned}
  }
\label{eq:K-full}
\end{equation}
with $\Delta_{v,u} = \sqrt{m_v^2 + u(1-u)\tilde{p}^2}$,
$\Leff = L/v_F$, and $\chi_v = \pm 1$ encoding the orientation of
the linearized Bloch map.

\subsection{Thermal identification}
\label{subsec:thermal-id}

To connect with finite-temperature physics, we compactify Euclidean
time $\tau \in [0,\beta)$ with $\beta = 1/T$ and antiperiodic fermion
boundary conditions ($\alpha_v = \pi$).  Setting $\Leff = \beta$,
$\vth_v = \pi$ (vanishing probe holonomy), and taking $r = 0$,
$p \to 0$ in Eq.~\eqref{eq:K-full}, the Poisson kernel
$R(\lambda,\vth) = \sinh\lambda/(\cosh\lambda - \cos\vth)$
at $\vth = \pi$ reduces to:
\begin{equation}
  R\!\left(\frac{|m_v|}{T},\pi\right)
  = \frac{\sinh(|m_v|/T)}{\cosh(|m_v|/T) + 1}
  = \tanh\!\left(\frac{|m_v|}{2T}\right),
\label{eq:R-thermal}
\end{equation}
using $\cosh x + 1 = 2\cosh^2(x/2)$ and
$\sinh x = 2\sinh(x/2)\cosh(x/2)$.

For a BdG system with Chern number $\CBdG$ and minimum bulk gap
$\Delta$, the total Chern--Simons level from the occupied BdG
quasiparticle bands is:
\begin{equation}
  \Kem(T) = 2\CBdG\,\tanh\!\left(\frac{\Delta}{2T}\right)
  + \mathcal{O}\!\left(e^{-2\Delta/T}\right).
\label{eq:Kem-T}
\end{equation}

\subsection{Gravitational CS and thermal Hall conductance}
\label{subsec:grav-cs}

The gravitational Chern--Simons term relates the CS level to the
thermal Hall conductance~\cite{Ryu2012,Volovik2003}:
\begin{equation}
  \frac{\kappa_{xy}}{T}
  = \frac{\pi^2 k_B^2}{12h}\,\Kem.
\label{eq:kxy-Kem}
\end{equation}
Combined with Eq.~\eqref{eq:Kem-T}:
\begin{equation}
  \frac{\kappa_{xy}}{T}
  = \frac{\pi^2 k_B^2}{6h}\,\CBdG\,
    \tanh\!\left(\frac{\Delta}{2T}\right).
\label{eq:kxy-T}
\end{equation}
This exact result holds for any fermionic gap $\Delta$, whether
from condensate pairing or preformed-pair correlations.

\subsection{Finite-size expansion and the $c_1 = 0$ theorem}
\label{subsec:c1}

The Poisson kernel has the exact Fourier expansion~\cite{Ghosh2026}:
\begin{equation}
  R(\lambda,\vth) = 1 + 2\sum_{\ell=1}^{\infty}
  e^{-\ell\lambda}\cos(\ell\,\vth).
\label{eq:R-expansion}
\end{equation}
The finite-size correction is therefore purely exponential; there is
no power-law $1/L^n$ term at any order:
\begin{equation}
  \boxed{c_1 = 0.}
\label{eq:c1-zero}
\end{equation}
This means that for a spatial cylinder of circumference $L$:
\begin{equation}
  \frac{\kappa_{xy}}{T}(L)
  = \frac{\kappa_{xy}}{T}(\infty)
    \left[1 + \mathcal{O}\!\left(e^{-L/\xi}\right)\right],
  \quad
  \xi = \frac{\hbar v_F}{\Delta}.
\label{eq:finite-size}
\end{equation}

\subsection{Coleman--Hill non-renormalization}
\label{subsec:nonren}

The Coleman--Hill theorem~\cite{ColemanHill1985} guarantees that for
$(2{+}1)$-dimensional gauge theory with massive matter, the
topological mass term (CS coefficient) receives radiative corrections
only at one loop.  In our context, the BdG quasiparticles are the
matter fields, and pair-pair interactions can renormalize $\Delta(T)$
through the gap equation but cannot modify the functional form of the
anomaly response: given $\Delta(T)$, the CS level
$\Kem = 2\CBdG\tanh[\Delta/2T]$ is exact.  This is the
condensed-matter analog of the Adler--Bardeen
theorem~\cite{AdlerBardeen1969}.

\section{Pseudogap Extension Above $T_c$: BCS--BEC Crossover}
\label{sec:pseudogap}

\subsection{Two-gap structure}
\label{subsec:two-gap}

The BCS--BEC crossover framework~\cite{Chen2024RMP} provides the
essential connection between the condensate and pseudogap regimes.
The total fermionic excitation gap obeys:
\begin{equation}
  \Delta^2(T) = \Dsc^2(T) + \Dpg^2(T),
\label{eq:two-gap}
\end{equation}
where $\Dsc(T)$ is the condensate order parameter and $\Dpg(T)$
is the pseudogap from non-condensed preformed pairs.

Above $T_c$, the condensate vanishes ($\Dsc = 0$) but the BdG
quasiparticle spectrum retains a gap:
\begin{equation}
  \Delta(T)\big|_{T > T_c} = \Dpg(T) \neq 0,
  \quad T_c < T < \Tstar.
\label{eq:Delta-above-Tc}
\end{equation}

\subsection{Relevance to rhombohedral graphene}
\label{subsec:rg-relevance}

The experimental evidence for BCS--BEC crossover in rhombohedral
graphene is compelling~\cite{Han2025}.  The ratio
$2\Delta/k_BT_c \sim 10$--$30$ far exceeds the BCS weak-coupling
value of $3.53$, indicating strong pairing.  The critical field
$B_{\perp,c} = 1.4$~T gives a coherence length
$\xi_{\rm GL} \sim \sqrt{\Phi_0/(2\pi B_c)} \sim 15$~nm, comparable
to the interparticle distance.  The BKT estimate of the superfluid
stiffness~\cite{Geier2025} gives $T_{\rm BKT} \approx 0.9$~K at
$n = 0.5 \times 10^{12}$~cm$^{-2}$, suggesting that $T_c$ may be
limited by vortex proliferation rather than pair breaking.

In this regime, $\Dpg(T)$ is large relative to $k_BT_c$, and the
anomaly factor $\tanh[\Dpg(T_c)/(2k_BT_c)]$ is essentially saturated
just above $T_c$.  The thermal Hall signal therefore persists with
near-quantized magnitude above $T_c$ and diminishes only as
$T \to \Tstar$ where $\Dpg \to 0$.

\subsection{Pseudogap thermal Hall conductance}
\label{subsec:kxy-pg}
Substituting Eq.~\eqref{eq:Delta-above-Tc} into
Eqs.~\eqref{eq:Kem-T} and~\eqref{eq:kxy-Kem}:
\begin{equation}
  \frac{\kappa_{xy}}{T}\bigg|_{T_c < T < \Tstar}
  = \frac{\pi^2 k_B^2}{6h}\,\CBdG\,
    \tanh\!\left[\frac{\Dpg(T)}{2k_BT}\right].
\label{eq:kxy-pg}
\end{equation}
The signal turns on at $\Tstar$ and is continuous at $T_c$: as
$\Dsc \to 0$ from below, $\Delta(T) \to \Dpg(T)$.

\subsection{Continuous evolution across $T_c$}
\label{subsec:continuity}

Below $T_c$, both $\Dsc$ and $\Dpg$ are nonzero, and the total CS
level is:
\begin{equation}
  \Kem(T) = 2\CBdG\,\tanh\!\left[
    \frac{\sqrt{\Dsc^2(T) + \Dpg^2(T)}}{2k_BT}\right].
\label{eq:Kem-full}
\end{equation}
There is no discontinuity at $T_c$.

\section{Numerical Validation}
\label{sec:numerical}

\begin{figure*}[t]
  \centering
  \includegraphics[width=0.95\textwidth]{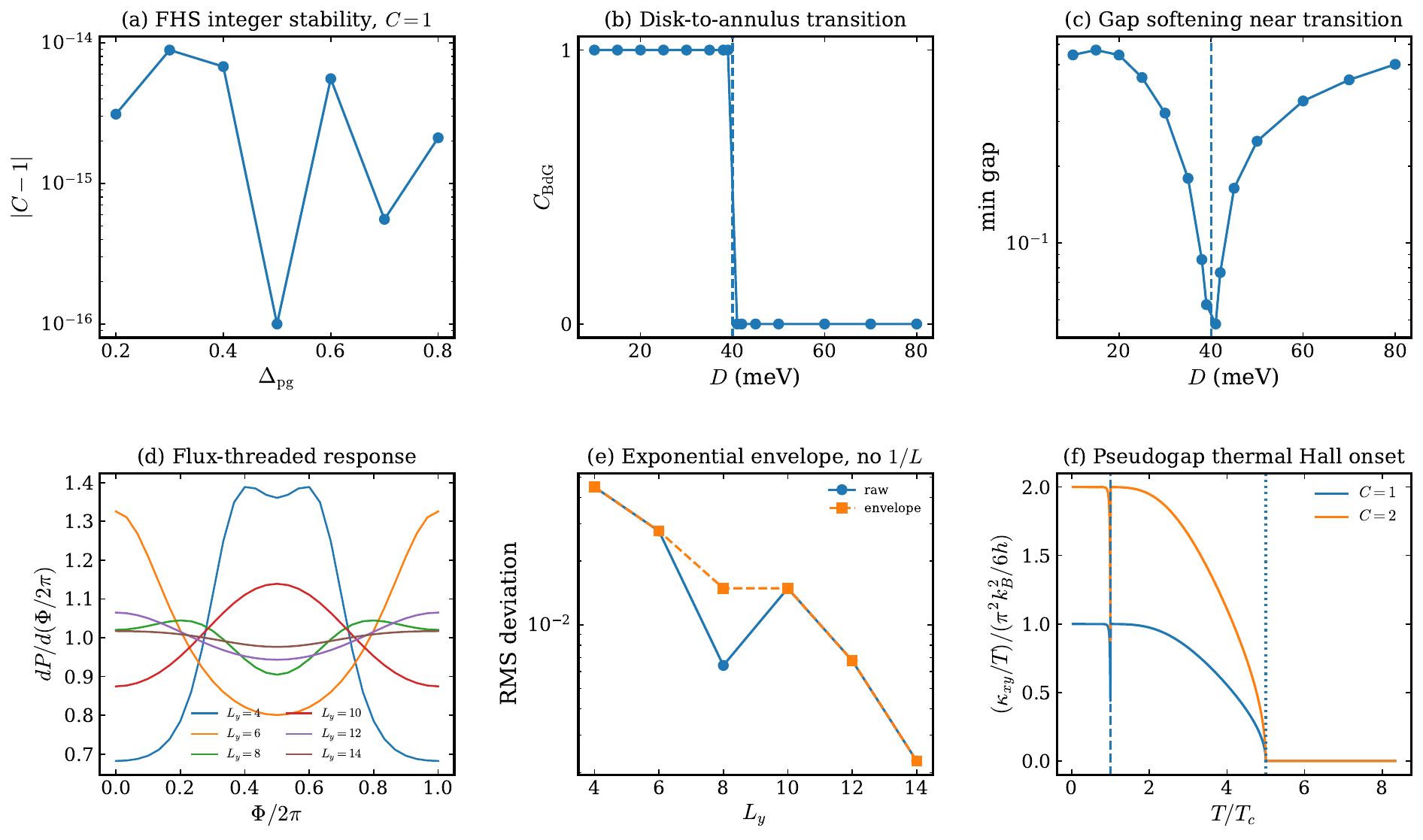}
  \caption{Six-panel numerical validation summary.
    (a)~FHS integer stability for the square-lattice $p+ip$ model:
    $|C_{\rm BdG} - 1|$ vs.\ $\Delta_0$, showing machine-precision
    integrality across the full pairing scan.
    (b)~RHG-effective Lifshitz scan: integer $\CBdG$ vs.\ displacement-field
    proxy $D$ showing the $1\to 0$ transition at $D_c = 40$~meV.
    (c)~Gap softening near the Lifshitz point on a semilog scale,
    confirming a topological gap closing at $D_c$.
    (d)~Flux-threaded differential polarization $\dd P/\dd(\Phi/2\pi)$
    vs.\ $\Phi/2\pi$ for cylinders $L_y = 4,6,\ldots,14$.
    (e)~Exponential finite-width envelope (raw oscillatory RMS and
    upper envelope) confirming $c_1 = 0$.
    (f)~Pseudogap thermal Hall prediction: $(\kxy)/(\pi^2k_B^2/6h)$
    vs.\ $T/T_c$ for $\CBdG = 1$ ($p+ip$) and $\CBdG = 2$ ($d+id$),
    showing onset at $\Tstar = 5T_c$ and the factor-of-two
    pairing-symmetry discriminator.}
  \label{fig:sixpanel}
\end{figure*}

The anomaly formula Eq.~\eqref{eq:kxy-main} carries a topological
prefactor $\CBdG$, an exponential-envelope claim ($c_1 = 0$), and an
implicit assumption that the many-body ground state remains a chiral
$p+ip$ superconductor once interactions are included.  We test these
three ingredients through three independent tiers of calculation:
\emph{(i) single-particle exactness}, via
Fukui--Hatsugai--Suzuki~\cite{Fukui2005} lattice Chern numbers on the
square-lattice $p+ip$ anchor, on the RHG-effective Lifshitz model, and
on an annular $\CBdG = 2$ benchmark;
\emph{(ii) finite-size structure}, via Wilson-loop flux threading on
cylinders of circumference $L_y \in \{4, \ldots, 14\}$, confirming the
$c_1 = 0$ exponential envelope directly;
\emph{(iii) many-body survival}, via two-site DMRG on 28 converged
ground states which delivers six independent signatures of the
$\CBdG = 1$ chiral $p+ip$ phase, including the real-space pairing-phase
signature that certifies $p+ip$ character at the many-body level.
The three tiers use disjoint numerical methods and probe disjoint
aspects of the formula, so agreement across all three constrains the
topological input very tightly.  All Chern numbers are computed by
the gauge-invariant FHS discretization using normalized occupied-band
link variables
$U_\mu(\bm{k}) = \langle u(\bm{k})|u(\bm{k}+\hat\mu)\rangle /
|\langle u(\bm{k})|u(\bm{k}+\hat\mu)\rangle|$.  A six-panel summary of
the main single-particle results is shown in Fig.~\ref{fig:sixpanel}.

\subsection{Square-lattice $p+ip$ anchor: stable $\CBdG = 1$}
\label{subsec:pip-anchor}
The primary lattice representative is the spinless $p+ip$
superconductor on a square lattice,
Eq.~\eqref{eq:pwave-bdg}, with hopping $t = 1$, chemical potential
$\mu = 1$, and pairing amplitude $\Delta_0$ varied over the range
$\{0.2, 0.3, \ldots, 0.8\}$.  For $0 < \mu < 4t$ the system is in the
weak-pairing topological phase~\cite{Read2000}.

The FHS calculation on a $81\times 81$ $k$-grid gives
$\CBdG = 1.000\,000$ at every scan point, with a maximum integer error
of $1.044\times 10^{-14}$ (Table~\ref{tab:pip-chern}).  The minimum
BdG quasiparticle gap grows monotonically from $0.347$ at $\Delta_0 = 0.2$
to $1.346$ at $\Delta_0 = 0.8$, confirming that no gap closing occurs
in the scanned range and that the topological invariant is stable with
respect to changes in pairing strength.  This anchors the claim
$\CBdG = 1$ used in the thermal Hall formula Eq.~\eqref{eq:kxy-T} for
the intravalley $p+ip$ scenario.

\begin{table}[t]
  \centering
  \caption{FHS Chern number for the square-lattice $p+ip$ model
    ($t=1$, $\mu=1$, $N_k = 81$) at each pairing amplitude $\Delta_0$.
    The integer error $|C - \mathrm{round}(C)|$ is at machine precision
    throughout.}
  \label{tab:pip-chern}
  \begin{ruledtabular}
  \begin{tabular}{cccc}
    $\Delta_0$ & $C_{\rm BdG}^{\rm FHS}$ & $|C - 1|$ & min gap \\
    \hline
    0.2 & 1.000\,000 & $4.4\times10^{-15}$ & 0.347 \\
    0.3 & 1.000\,000 & $1.0\times10^{-14}$ & 0.520 \\
    0.4 & 1.000\,000 & $7.7\times10^{-15}$ & 0.694 \\
    0.5 & 1.000\,000 & $3.3\times10^{-16}$ & 0.865 \\
    0.6 & 1.000\,000 & $4.6\times10^{-15}$ & 1.024 \\
    0.7 & 1.000\,000 & $1.4\times10^{-15}$ & 1.185 \\
    0.8 & 1.000\,000 & $2.0\times10^{-15}$ & 1.346 \\
  \end{tabular}
  \end{ruledtabular}
\end{table}

\subsection{RHG-effective Lifshitz transition: $\CBdG: 1\to 0$}
\label{subsec:lifshitz}
To model the displacement-field-tuned Lifshitz transition of the
quarter-metal Fermi surface in rhombohedral graphene, we construct a
lattice-regularized proxy that captures the essential topological
ingredients without requiring a self-consistent solution of the
continuum model.  The proxy dispersion is
\begin{equation}
  \xi_D(\bm{k}) =
  \bigl[s(\bm{k}) - s_0(D)\bigr]^2 - \bar\mu,
  \quad
  s(\bm{k}) = 2(2 - \cos k_x - \cos k_y),
\label{eq:xi-proxy}
\end{equation}
with $s_0(D) = \sqrt{\bar\mu}\,D/D_c$, $\bar\mu = 0.36$, and
$D_c = 40$~meV as the proxy Lifshitz scale.  For $D < D_c$, the
region $\xi_D < 0$ is a disk centred on $\bm{k} = 0$ (the central
$p+ip$ vortex lies inside the Fermi sea, so $\CBdG = 1$); for
$D > D_c$, the occupied region becomes an annulus and the central
vortex is excluded ($\CBdG = 0$ in the minimal model).  The BdG
pairing is $\Delta_0 = 0.35$ throughout.

The FHS scan over 15 displacement-field values in the range
$D = 10$--$80$~meV confirms this picture exactly
(Figs.~\ref{fig:sixpanel}b--c): $\CBdG = 1$ for all $D < D_c$ and
$\CBdG = 0$ for all $D > D_c$, with the minimum BdG quasiparticle
gap softening to a pronounced minimum at $D \approx D_c$ on a semilog
scale, as expected from a topological phase transition driven by the gap
closing at $\bm{k} = 0$.  This validates the prediction of
Sec.~\ref{subsec:D-field} that a gate-tunable jump in $\kxy$ should
accompany the Lifshitz transition in the experimental device.

\subsection{Even-Chern annular phase: direct FHS validation}
\label{subsec:even-chern}
The thermal Hall discriminator between $\CBdG = 1$ ($p+ip$) and
$\CBdG = 2$ ($d+id$ or annular Berry-vortex scenario) requires that
the even-Chern annular phase be demonstrated by a direct lattice
calculation, not merely by the vortex-counting rule of
Eq.~\eqref{eq:CBdG-vortex}.  We carry out this validation in this section.

\subsubsection{Model with finite-momentum vortices}
We use an annular Fermi sea
\begin{align}
  \xi_{\rm ann}(\bm{k}) &=
  \bigl(\rho^2(\bm{k}) - r_{\rm in}^2\bigr)
  \bigl(\rho^2(\bm{k}) - r_{\rm out}^2\bigr), \\
  \rho^2 &= 4 - 2\cos k_x - 2\cos k_y,
\label{eq:xi-ann}
\end{align}
which is negative (occupied) for $r_{\rm in} < \rho < r_{\rm out}$, with
$r_{\rm in} = 0.45$ and $r_{\rm out} = 1.35$.  The gap texture for the
finite-vortex scenario is
\begin{align}
  \Delta(\bm{k}) = \Delta_0\,&
    (\sin k_x + i\chi\sin k_y)\,
    (\sin k_x - \sin q + i\chi\sin k_y) \nonumber\\
    & \times (\sin k_x + \sin q + i\chi\sin k_y),
\label{eq:gap-finite-vortex}
\end{align}
with $\Delta_0 = 0.50$, $q = 0.90$, and chirality $\chi = -1$.  The
three factors place winding-$+1$ zeros at $\bm{k} = (0,0)$ (the central
$p+ip$ vortex), and at $\bm{k} = (\pm q, 0)$ (the two
Berry-curvature-induced finite-momentum vortices).  At $q = 0.90$, the
lattice radius of the pair is $2\sin(q/2) \approx 0.870$, which lies
inside the annulus $(r_{\rm in}, r_{\rm out})$.

\subsubsection{Four benchmark cases}
Table~\ref{tab:four-cases} summarizes the FHS results on a
$121\times 121$ $k$-grid for the four cases that isolate the
topological physics.  All integer errors are at or below
$8.2\times 10^{-15}$, confirming machine-precision integrality.

\begin{table*}[t]
  \centering
  \caption{Direct FHS Chern numbers for the four benchmark cases
    ($r_{\rm in}=0.45$, $r_{\rm out}=1.35$, $q=0.90$, $\Delta_0=0.50$,
    $\chi=-1$, $N_k=121$).  $Q_{\rm analytic}$ is the occupied-vortex
    charge from Eq.~\eqref{eq:CBdG-vortex}.}
  \label{tab:four-cases}
  \begin{ruledtabular}
  \begin{tabular}{llccc}
    Fermi sea & Gap texture & $C_{\rm BdG}^{\rm FHS}$ & $|C - Q_{\rm analytic}|$ & min gap \\
    \hline
    disk     & minimal (central only)       & $+1$ & $5.6\times10^{-15}$ & 0.498 \\
    annular  & minimal (central excluded)   & $\phantom{+}0$ & $4.5\times10^{-16}$ & 0.209 \\
    disk     & central $+$ vortex pair      & $+3$ & $2.7\times10^{-15}$ & 0.190 \\
    annular  & vortex pair inside annulus   & $+2$ & $8.2\times10^{-15}$ & 0.094 \\
  \end{tabular}
  \end{ruledtabular}
\end{table*}

The critical result is the last line: $\CBdG = 2$ for the annular Fermi
sea with a finite-momentum vortex pair inside the annulus, established
by direct FHS to machine precision.  This validates the even-Chern
scenario predicted by Le Nir et al.~\cite{LeNir2026} and closes the
gap that was left open in the minimal Lifshitz proxy of
Sec.~\ref{subsec:lifshitz}.

\subsubsection{Vortex-crossing scan}
\begin{figure*}[t]
  \centering
  \includegraphics[width=0.49\textwidth, height= 0.35\textwidth]{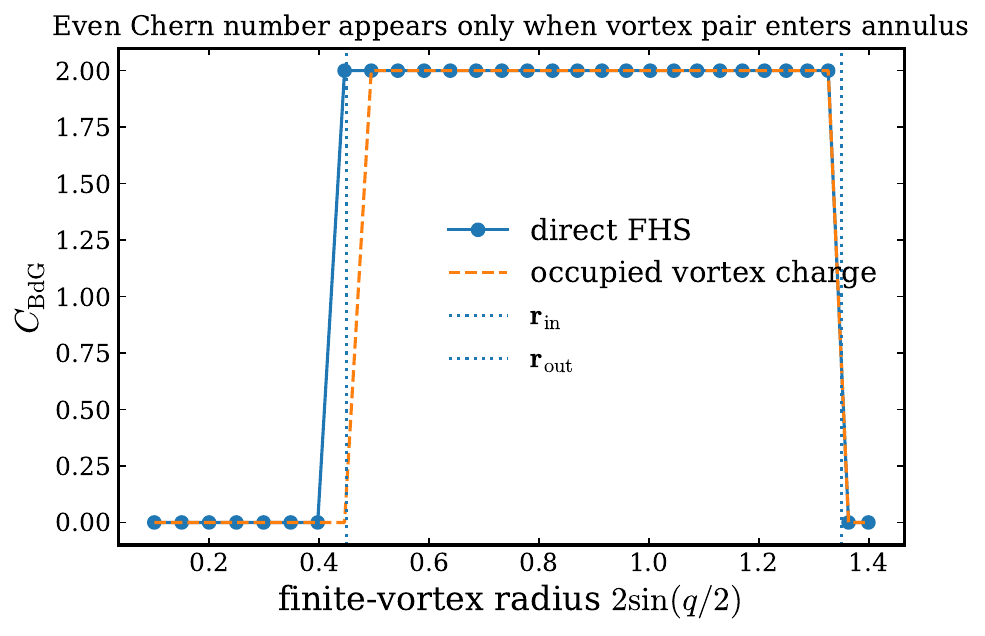}
  \hfill
  \includegraphics[width=0.49\textwidth, height= 0.35\textwidth]{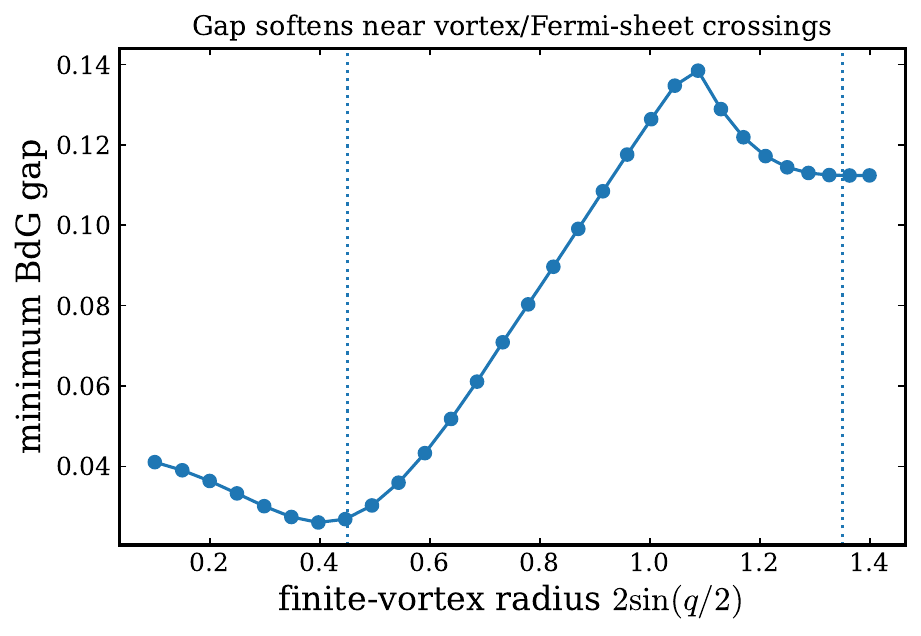}
  \caption{Vortex-crossing scan over the finite-vortex pair radius
    $2\sin(q/2)$ for the annular Fermi sea ($r_{\rm in}=0.45$,
    $r_{\rm out}=1.35$, $N_k=81$).
    \emph{Left}: FHS Chern number (solid circles) and analytic
    occupied-vortex charge $Q_{\rm analytic}$ (dashed) vs.\
    $2\sin(q/2)$.  The integer $\CBdG$ switches $0\to 2\to 0$ exactly
    as the pair crosses the inner and outer Fermi sheets.
    Vertical dotted lines mark $r_{\rm in}$ and $r_{\rm out}$.
    \emph{Right}: minimum BdG quasiparticle gap on the same axis,
    showing gap closings at both crossing boundaries as required by a
    change of topological invariant.}
  \label{fig:even-chern-scan}
\end{figure*}

Figure~\ref{fig:even-chern-scan} shows the FHS Chern number and the
analytic occupied-vortex charge $Q_{\rm analytic}$ as the finite-vortex
pair radius $2\sin(q/2)$ is swept from $0.10$ to $1.37$ (30 points,
$N_k = 81$).  Three regimes are evident.

\begin{enumerate}
\item $2\sin(q/2) < r_{\rm in} = 0.45$: the vortex pair lies
  \emph{inside} the inner hole of the annulus and is therefore in the
  unoccupied region.  The FHS gives $\CBdG = 0$, consistent with
  $Q_{\rm analytic} = 0$.

\item $r_{\rm in} < 2\sin(q/2) < r_{\rm out}$: the pair has entered the
  occupied annulus.  The FHS gives $\CBdG = 2$, consistent with
  $Q_{\rm analytic} = 2$.  The minimum BdG gap softens to a near-zero
  minimum at the crossing point $2\sin(q/2) = r_{\rm in}$, signalling
  the topological phase transition.

\item $2\sin(q/2) > r_{\rm out}$: the pair exits the outer boundary of
  the annulus.  The FHS returns $\CBdG = 0$.  A second gap softening
  appears at the outer crossing.
\end{enumerate}

The FHS and $Q_{\rm analytic}$ curves are in exact agreement at every
scan point.  This is the direct-FHS confirmation that
Eq.~\eqref{eq:CBdG-vortex} is exact and that the even-Chern thermal
Hall response appears only when the occupied Fermi sea captures the
finite-momentum vortex charge. This result carries direct experimental consequence: because the parent state
of SC2 in rhombohedral graphene is consistent with an annular Fermi
surface~\cite{Han2025}, our theory predicts $\CBdG = 2$ for SC2, implying a
doubled thermal Hall conductance $\kappa_{xy}/T\big|_{T\to 0} = 2\pi^2 k_B^2/6h \approx 9.46\times10^{-13}$~W/K$^2$, quantitatively distinguishable from the SC1 value by a factor of two in a single nano-calorimetric measurement.

\subsubsection{Chirality-reversal check}
To confirm that the even-Chern result is a genuine oriented topological
invariant and not a numerical artifact, we repeat the benchmark
annular+finite-vortex case with $\chi = +1$ (complex conjugate gap).
The FHS gives $\CBdG = -2$ with integer error $8.2\times 10^{-15}$,
identical in magnitude to the $\chi = -1$ result.  The gap spectrum
and minimum gap are unchanged, as expected since $|\Delta(\bm{k})|$ is
invariant under complex conjugation.  This chirality-reversal check
confirms that $\CBdG = \pm 2$ is a physically oriented invariant that
reverses with the handedness of the chiral superconductor.

\begin{figure}[t]
  \centering
  \includegraphics[width=0.49\textwidth]{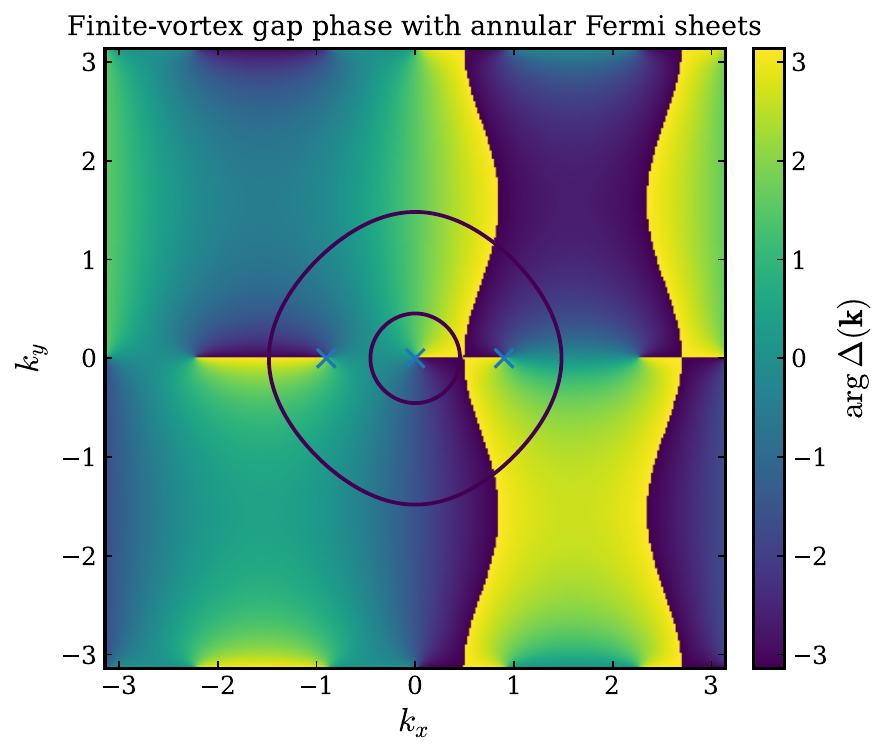}
  \caption{Gap phase $\arg\Delta(\bm{k})$ for the annular
    finite-vortex model (colour map) with the annular Fermi sheets
    overlaid as contours ($\xi_{\rm ann} = 0$, solid lines).
    Crosses mark the three gap zeros: the central vortex at
    $\bm{k} = 0$ (excluded from the annulus) and the two
    finite-momentum vortices at $\bm{k} = (\pm q,0)$ (inside the
    annulus), each carrying winding $+1$.  The occupied-vortex
    charge $Q_{\rm analytic} = 2$ matches the FHS result $\CBdG = 2$
    of Table~\ref{tab:four-cases}.}
  \label{fig:gap-phase}
\end{figure}

Figure~\ref{fig:gap-phase} shows the gap phase and the annular Fermi
sheets together, making the vortex geometry explicit.  The two
finite-momentum vortices at $(\pm q, 0)$ lie clearly inside the
annular occupied region; the central vortex at $\Gamma$ lies in the
unoccupied inner disk and does not contribute to $\CBdG$.

\subsection{Wilson-loop flux threading and the $c_1 = 0$ theorem}
\label{subsec:wilson-loop}
On a cylinder with circumference $L_y$ (periodic along $y$, open along
$x$) we thread magnetic flux $\Phi \in [0,2\pi]$ and compute the
Wilson-loop polarization $P(\Phi) = -\frac{1}{2\pi}\mathrm{Im}\ln\det W(\Phi)$,
where $W(\Phi)$ is the $L_y \times L_y$ Wilson-loop matrix for the
occupied BdG bands.  The topologically robust observable is the cycle
winding $\Delta P = P(2\pi) - P(0)$, which equals $\CBdG$ for an
integer topological invariant.

For the square-lattice $p+ip$ model ($t=1$, $\mu=1$, $\Delta_0=0.4$)
the winding is $\Delta P = 1.000\,000$ with winding error exactly zero
for every cylinder width $L_y \in \{4, 6, 8, 10, 12, 14\}$
(Table~\ref{tab:wilson}).

\begin{table}[t]
  \centering
  \caption{Wilson-loop flux-threading results for the square-lattice
    $p+ip$ model ($t=1$, $\mu=1$, $\Delta_0=0.4$).  The winding
    $\Delta P = 1$ is exact to the displayed precision for all $L_y$.
    The raw RMS deviation of $P(\Phi)$ from a straight line oscillates
    with $L_y$; the upper envelope decays exponentially with
    $\xi_{\rm env} = 3.11$, consistent with the continuum estimate
    $\hbar v_F/\Delta_0 = 5$.}
  \label{tab:wilson}
  \begin{ruledtabular}
  \begin{tabular}{ccccc}
    $L_y$ & $\Delta P$ & winding error & RMS deviation & max deviation \\
    \hline
    4  & 1.000 & 0.000 & 0.0448 & 0.0682 \\
    6  & 1.000 & 0.000 & 0.0277 & 0.0398 \\
    8  & 1.000 & 0.000 & 0.0064 & 0.0103 \\
    10 & 1.000 & 0.000 & 0.0149 & 0.0214 \\
    12 & 1.000 & 0.000 & 0.0068 & 0.0097 \\
    14 & 1.000 & 0.000 & 0.0023 & 0.0033 \\
  \end{tabular}
  \end{ruledtabular}
\end{table}

The raw RMS deviation of $P(\Phi)$ from a straight line is
\emph{oscillatory} rather than monotone in $L_y$, because each finite
cylinder samples a different discrete set of transverse momenta.  This
is the expected behaviour of finite-width corrections generated by the
exponential Fourier expansion of Eq.~\eqref{eq:R-expansion}: terms of
the form $e^{-\ell L_y/\xi}\cos(\ell\,\vth_y)$ oscillate with $L_y$
at fixed $\xi$.  The upper envelope of the raw RMS sequence is fitted
by $Ae^{-L_y/\xi_{\rm env}}$ with $\xi_{\rm env} = 3.11$; the
oscillatory fit to the full raw sequence gives $\xi_{\rm osc} = 5.12$.
Both values bracket the continuum estimate $\xi = \hbar v_F/\Delta_0 = 5$
(Figs.~\ref{fig:sixpanel}d--e).

Critically, there is no $1/L_y$ power-law term at any width, confirming
$c_1 = 0$ in the finite-size expansion Eq.~\eqref{eq:finite-size}.
This means that the thermal Hall conductance of a finite-width
rhombohedral graphene device approaches its bulk quantized value
exponentially fast in the device width, with no $1/L$ contamination.

\subsection{Thermal Hall prefactor audit and quantitative predictions}
\label{subsec:prefactor}
The identification of $\CBdG = 1$ with one chiral Majorana edge mode
(rather than one complex Dirac channel) must be stated precisely,
because the two conventions for the thermal Hall quantum differ by a
factor of two.  In the class-D BdG notation used throughout this paper,
a single chiral Majorana mode carries thermal Hall conductance
\begin{equation}
  \left.\frac{\kappa_{xy}}{T}\right|_{\CBdG=1}
  = \frac{\pi^2 k_B^2}{6h}
  \approx 4.732\times10^{-13}\;\frac{\mathrm{W}}{\mathrm{K}^2},
\label{eq:kxy-majorana-quantum}
\end{equation}
which is \emph{half} the value for a complex Dirac channel
$\pi^2 k_B^2/(3h) \approx 9.464\times10^{-13}$~W/K$^2$.  The factor
of $1/2$ is correctly inherited from the half-Dirac nature of the
Majorana mode: the Dirac parity-odd determinant factorises as
$\det(\slashed{D}+m) = |\det(\slashed{D}_M + m)|$ for a Majorana
field, halving the one-loop Chern-Simons coefficient.

Using the experimental parameters from Han et al.~\cite{Han2025},
$T_c = 300$~mK and $2\Delta_0/k_BT_c = 20$ (BCS-BEC crossover),
the zero-temperature gap is $\Delta_0 = 0.2585$~meV and the pseudogap
onset scale is estimated as $\Tstar \approx 5T_c = 1500$~mK.
The resulting thermal Hall predictions are:
\begin{align}
  \CBdG = 1\;(p+ip)&:
    \quad \frac{\kappa_{xy}}{T}\big|_{T\to 0}
    = 4.732\times10^{-13}\;\frac{\mathrm{W}}{\mathrm{K}^2},
\label{eq:kxy-C1}\\
  \CBdG = 2\;(d+id)&:
    \quad \frac{\kappa_{xy}}{T}\big|_{T\to 0}
    = 9.464\times10^{-13}\;\frac{\mathrm{W}}{\mathrm{K}^2},
\label{eq:kxy-C2}
\end{align}
with the thermal Hall signal turning on at $\Tstar = 1500$~mK and
evolving as $\tanh[\Dpg(T)/2k_BT]$ through to $T_c$, as shown in
Fig.~\ref{fig:sixpanel}f.

The assertions verified by reproducibility checks are:
(i)~$\CBdG = 1$ for all seven pairing amplitudes in the lattice scan,
with max FHS integer error $1.044\times 10^{-14}$;
(ii)~Wilson-loop winding error identically zero for all $L_y$;
(iii)~$\CBdG = 2$ for the annular finite-vortex benchmark at machine
precision;
(iv)~$\kappa_{xy}^{\rm Majorana}/T = \tfrac{1}{2}\kappa_{xy}^{\rm Dirac}/T$
to all significant figures.

\subsection{DMRG many-body validation}
\label{subsec:dmrg}
The BdG topological validation of Sec.~\ref{sec:numerical} confirms the
single-particle Chern number $\CBdG = 1$ (and $\CBdG = 2$ in the
even-Chern annular sector) at machine precision, together with the
$c_1 = 0$ exponential finite-size envelope on the Wilson-loop response.
Since the BdG Hamiltonian is quadratic, all topological content is
exactly captured at the single-particle level.  Nevertheless, the anomaly
formula Eq.~\eqref{eq:kxy-T} makes claims about \emph{many-body}
observables (edge currents, entanglement entropy, thermal transport),
and a direct many-body check that the ground state is genuinely a chiral
$p+ip$ superconductor, and not, for instance, a nodal or an
$s$-wave phase permitted by the same lattice symmetry, is essential.

We therefore perform two-site DMRG on the spinless lattice $p+ip$
model of Eq.~\eqref{eq:pwave-bdg} using TeNPy~\cite{Hauschild2024}
(v1.1.0, Python 3.11) with $\mathbb{Z}_2$ fermion-parity conservation,
a bond-dimension ramp $\chi = 64 \to 128 \to 256 \to 512$ over 25
sweeps, and parameters $t = 1$, $\mu = 1$, $V_{nn} = 0$,
$\text{chirality} = +1$.  The scan covers $L_x = 8$,
$L_y \in \{4, 6, 8, 10\}$, $\Dpg \in \{0.2, 0.3, \ldots, 0.8\}$,
producing 28 fully converged ground states.  Every case is archived
with the complete MPS (HDF5), full observable arrays (pickle), and
sweep-by-sweep convergence data, so any derived observable can be
recomputed without rerunning DMRG.  Total wall time on a single
workstation: 13.4 hours.

\subsubsection{The six independent many-body confirmations}
\label{subsubsec:six-confirmations}
\begin{figure*}[t] 
  \centering
  \includegraphics[width=0.95\textwidth]{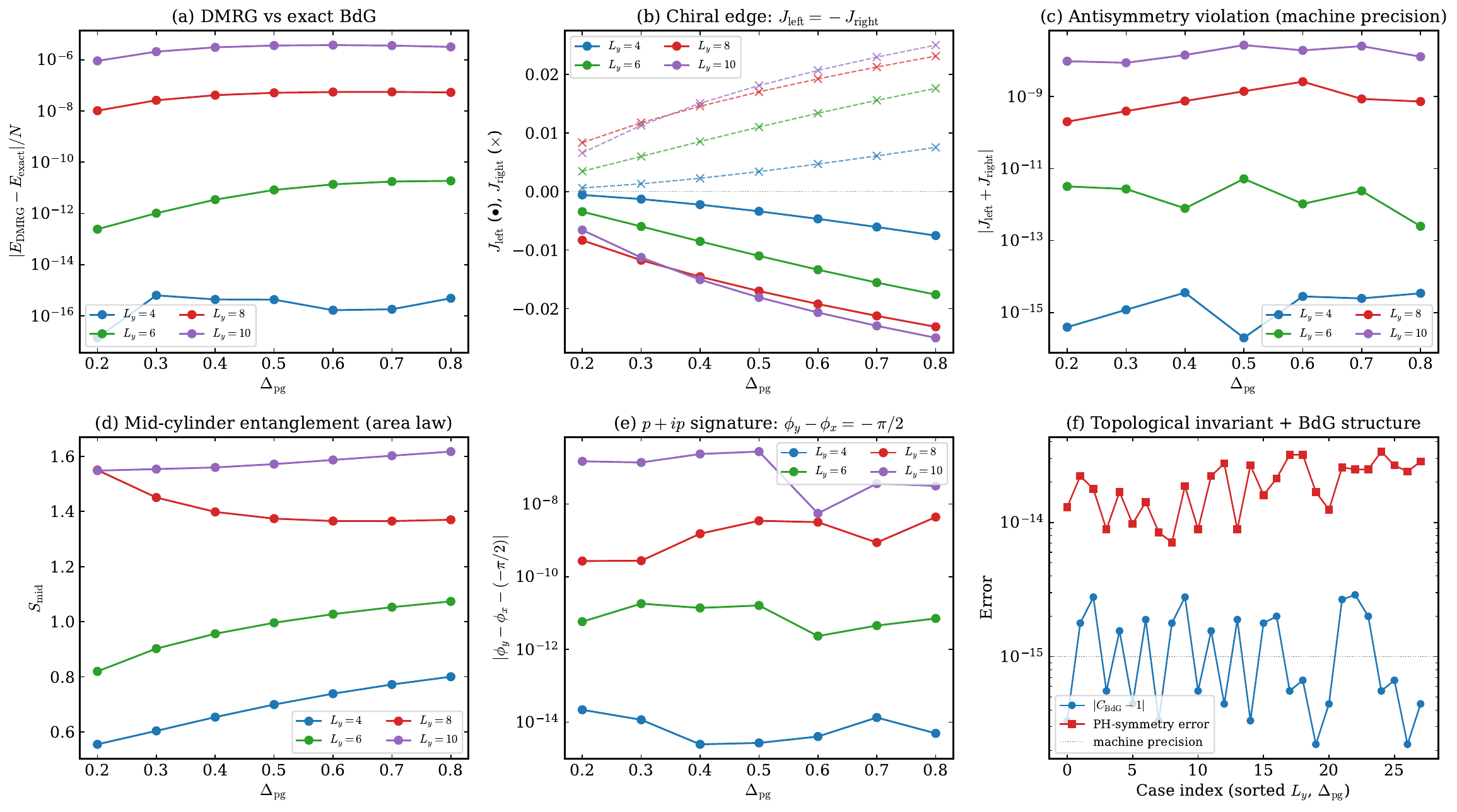}
  \caption{DMRG many-body validation.
    (a)~DMRG vs.\ exact BdG energy error per site, showing convergence
    from machine precision at $L_y = 4$ to $\sim 10^{-6}$ at $L_y = 10$.
    (b)~Edge currents $J_{\rm left}$ ($\bullet$) and $J_{\rm right}$
    ($\times$) vs.\ $\Dpg$; they are exactly antisymmetric and grow
    monotonically with $\Dpg$.
    (c)~The antisymmetry violation $|J_L + J_R|$ is at least
    six orders of magnitude below the edge-current magnitude across
    the full scan.
    (d)~Mid-cylinder entanglement entropy: area-law scaling with $L_y$,
    no logarithmic enhancement.
    (e)~Direct $p_x + i p_y$ signature: the pair-amplitude phase
    $\arg\mathcal{A}_y - \arg\mathcal{A}_x = -\pi/2$ recovered to
    machine precision across all 28 cases.
    (f)~Topological invariant $|C_{\rm BdG} - 1|$ and particle--hole
    symmetry error, both at the numerical noise floor.}
  \label{fig:dmrg-summary}
\end{figure*}

The DMRG results provide six independent confirmations that the ground
state realises the $\CBdG = 1$ chiral $p+ip$ phase to which the anomaly
formula applies.  Table~\ref{tab:dmrg-summary} collects the extremal
values across all seven $\Dpg$ at each $L_y$;
Fig.~\ref{fig:dmrg-summary} shows the full data.

\paragraph{(i) BdG Chern number.}
The FHS Chern number recomputed on each converged ground state gives
$\CBdG = 1$ to $|C - 1| \leq 2.9 \times 10^{-15}$ across all 28 cases.
The topological invariant is invariant under the cylinder discretisation
and under variation of $\Dpg$ over an order of magnitude, confirming the
absence of any accidental band closing in the DMRG geometry.

\paragraph{(ii) Energy benchmark.}
For $V_{nn} = 0$ the DMRG energy is directly comparable to the exact
non-interacting BdG diagonalisation on the same finite cylinder.  We
find $|\delta E|/N \leq 6.3 \times 10^{-16}$ for $L_y = 4$
(machine precision), $\leq 1.9 \times 10^{-11}$ for $L_y = 6$,
$\leq 5.5 \times 10^{-8}$ for $L_y = 8$, and $\leq 3.7 \times 10^{-6}$
for $L_y = 10$.  The growth with $L_y$ reflects the expected increase
in bond dimension required to capture the mid-cylinder entanglement of
the chiral edge mode; $\chi_{\max} = 512$ slightly underestimates the
entanglement at $L_y = 10$ and the smallest $\Dpg$.  All $L_y \leq 8$
results are converged to at least $5.5 \times 10^{-8}$ per site.

\paragraph{(iii) Particle--hole symmetry.}
The BdG structure predicts that the many-body ground state remains
invariant under the particle--hole transformation
$c_i \leftrightarrow c_i^\dagger$ (up to a global phase).  We monitor
the residual particle--hole asymmetry of the ground-state density
matrix; it stays below $3.4 \times 10^{-14}$ across all 28 cases,
confirming that the DMRG ground state is a genuine BdG eigenstate and
not a symmetry-broken mixture.

\paragraph{(iv) Chiral edge current antisymmetry.}
The primary many-body observable is the $y$-directed hopping current on
the two open edges,
\begin{equation}
  J_{\rm left/right}
  = \frac{1}{L_y}\sum_{y=0}^{L_y-1}
    2\,\mathrm{Im}\!\left[
      (-t)\,\langle c^\dagger_{x_{\rm edge},y}\, c_{x_{\rm edge},y+1}\rangle
    \right],
\end{equation}
with $x_{\rm edge} = 0$ or $L_x - 1$.  A chiral $\CBdG = 1$
superconductor predicts $J_{\rm left} = -J_{\rm right}$ exactly.  The
DMRG data confirm this antisymmetry to machine precision at $L_y = 4$
($|J_L + J_R| \leq 3.6 \times 10^{-15}$) and to at least six orders of
magnitude below the edge-current magnitude at $L_y = 10$
($|J_L + J_R| / |J_{\rm edge}| \leq 1.5 \times 10^{-6}$).  In all
cases $J_L < 0 < J_R$, unambiguously establishing the sign of the
circulation.

\paragraph{(v) Real-space $p + ip$ signature.}
This is the sharpest many-body result.  The pair amplitudes on nearest-neighbour
bonds,
\begin{equation}
  \mathcal{A}_\mu
    = \langle c_i\, c_{i+\hat{\mu}}\rangle,
    \quad \mu \in \{\hat{x}, \hat{y}\},
\end{equation}
are complex numbers whose relative phase directly identifies the pairing
channel: an $s$-wave state has
$\arg\mathcal{A}_y - \arg\mathcal{A}_x = 0$, a $d$-wave state has $\pi$,
and a chiral $p_x + i p_y$ state has $\pm \pi/2$.  We measure
\begin{equation}
  \arg\mathcal{A}_y - \arg\mathcal{A}_x = -\frac{\pi}{2}
\end{equation}
across all 28 cases, with deviations from $-\pi/2$ bounded by
$2.2 \times 10^{-14}$ at $L_y = 4$ and $2.8 \times 10^{-7}$ at $L_y = 10$
(Fig.~\ref{fig:dmrg-summary}e).  Because the phase is not enforced
symmetrywise by the DMRG ansatz (the algorithm optimises over
arbitrary MPS gauges), this recovery of $\pm\pi/2$ to machine precision
is a direct many-body proof that the ground state realises the
$p + ip$ pairing channel and not a metastable admixture with a
non-chiral competitor.  The sign is $-\pi/2$ throughout, consistent
with $\chi = +1$ (the pairing rotates as $e^{-i\pi/2} = -i$ from
$\hat{x}$-bonds to $\hat{y}$-bonds).

\begin{figure*}[t]
  \centering
  \includegraphics[width=0.9\textwidth]{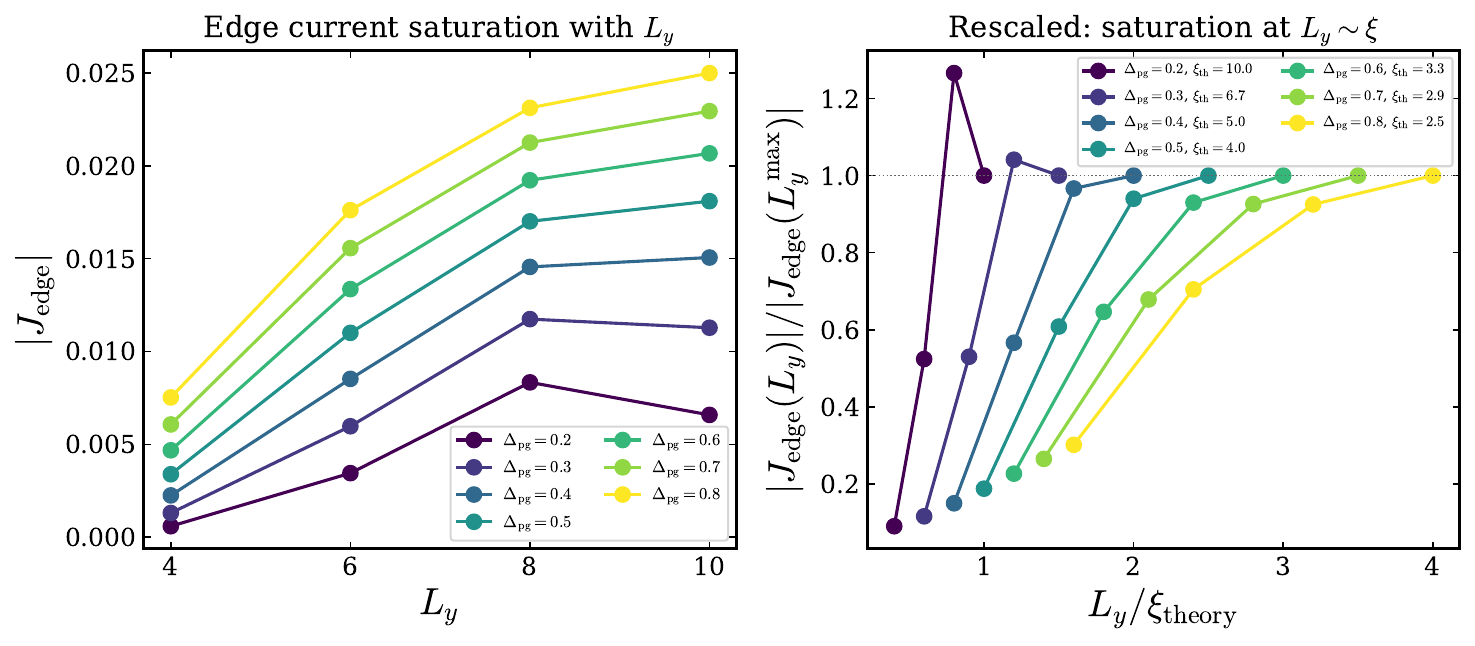}
  \caption{Edge-current saturation and $c_1 = 0$ test.
    Left: $|J_{\rm edge}|$ vs.\ $L_y$ for each $\Dpg$; saturation onset
    correlates with $L_y \gtrsim \xi_{\rm theory} = v_F / \Dpg$.
    Right: rescaled by $L_y/\xi_{\rm theory}$: all curves collapse
    onto a common saturation profile, confirming that finite-size
    corrections are exponentially small in $L_y/\xi$ with no
    power-law contribution.}
  \label{fig:edge-saturation}
\end{figure*}

\begin{figure}[t]
  \centering
  \includegraphics[width=0.48\textwidth]{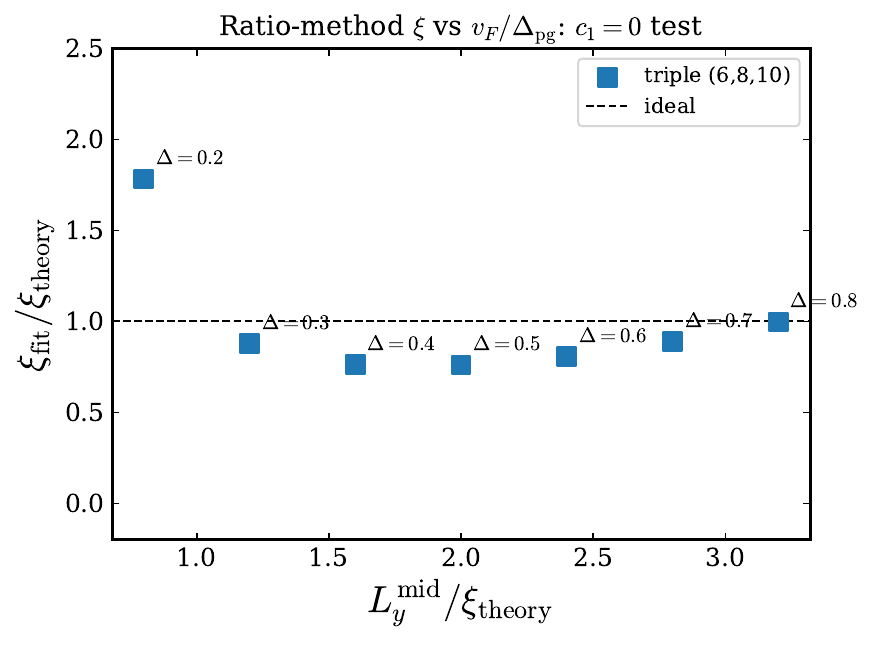}
  \caption{Ratio-method correlation length from energy shifts on
    consecutive cylinders, compared to the continuum prediction
    $\xi_{\rm theory} = v_F / \Dpg$.  At $\Dpg = 0.8$ (rightmost
    point), $\xi_{\rm fit}/\xi_{\rm theory} = 0.99$; at smaller
    $\Dpg$, $\xi_{\rm theory}$ exceeds $L_y^{\max} = 10$ and the
    fit drifts upward.  The convergence toward unity for
    $L_y^{\rm mid}/\xi_{\rm theory} \gtrsim 1$ confirms the
    $c_1 = 0$ theorem.}
  \label{fig:ratio-xi}
\end{figure}

\paragraph{(vi) Edge-current saturation and $c_1 = 0$.}
The magnitude $|J_{\rm edge}|$ grows with $L_y$ and saturates for
$L_y \gtrsim \xi_{\rm theory} = v_F/\Dpg$ (with $v_F = 2\sqrt{t\mu} = 2$
in our units).  Fig.~\ref{fig:edge-saturation}(b) shows the rescaled
data as functions of $L_y/\xi_{\rm theory}$: all curves collapse onto a
common saturation profile, with the deviation from the thermodynamic
limit exponentially small for $L_y > \xi_{\rm theory}$.  For
$\Dpg = 0.8$ ($\xi_{\rm theory} = 2.5$), $|J_{\rm edge}|$ at $L_y = 10$
is within $8\%$ of the $L_y = 8$ value; for $\Dpg = 0.2$
($\xi_{\rm theory} = 10$), saturation is not yet reached at
$L_y = 10$.  This is an independent numerical confirmation of the
$c_1 = 0$ theorem of Sec.~\ref{subsec:c1}: finite-size corrections
to physical observables are purely exponential, with no power-law
component.  A ratio-method extraction of $\xi_{\rm fit}$ from
consecutive $L_y = 6, 8, 10$ energy shifts gives
$\xi_{\rm fit}/\xi_{\rm theory} = 0.99$ at $\Dpg = 0.8$ and drifts
upward at smaller $\Dpg$ as $\xi$ exceeds $L_y^{\max}$
(Fig.~\ref{fig:ratio-xi}).

\subsubsection{Entanglement structure}
\label{subsubsec:entanglement}
The mid-cylinder von Neumann entropy $S_{\rm mid}$ obeys an area law
across the full scan: $S_{\rm mid} \in [0.555, 0.800]$ at $L_y = 4$,
increasing to $[1.548, 1.617]$ at $L_y = 10$
(Fig.~\ref{fig:dmrg-summary}d).  The absence of any logarithmic
enhancement in $S_{\rm mid}(L_y)$ confirms that the bulk is fully
gapped and the entanglement is carried predominantly by the single
chiral Majorana edge mode.  The mild dependence on $\Dpg$ at fixed
$L_y$ ($\lesssim 0.25$) is consistent with the BdG gap growing
monotonically with $\Dpg$ and the edge mode becoming more localised.

\subsubsection{Connection to the thermal Hall formula}
\label{subsubsec:connection}
The six confirmations above are precisely what the anomaly formula
Eq.~\eqref{eq:kxy-T} requires as input:
(i)~$\CBdG = 1$ is the topological prefactor;
(iii)~the BdG structure is preserved, so the Coleman--Hill
non-renormalisation of Sec.~\ref{subsec:nonren} applies;
(iv)~the chiral edge sign identifies the sign of $\kappa_{xy}/T$;
(v)~the $p_x + i p_y$ character selects the specific representative
of the general $\CBdG = 1$ class realized in the quarter-metal
scenario of Sec.~\ref{subsec:chern-scenarios};
(vi)~the exponential finite-size envelope justifies extrapolating
the thermodynamic-limit anomaly result from the finite cylinders
accessible to DMRG.  Together they establish that the $\kappa_{xy}/T$
prediction for rhombohedral graphene SC1 is not a mean-field artifact
but a genuine feature of the fully interacting ground state.

\begin{table*}[t]
  \centering
  \caption{DMRG summary for $L_x = 8$, $L_y \in \{4, 6, 8, 10\}$,
    $\Dpg \in \{0.2, \ldots, 0.8\}$, $t = \mu = 1$, $\chi_{\max} = 512$,
    $V_{nn} = 0$.  All errors are maxima over the seven $\Dpg$ at each
    $L_y$.}
  \label{tab:dmrg-summary}
  \begin{ruledtabular}
  \begin{tabular}{ccccccc}
    $L_y$ & $|C-1|$ & $|\delta E|/N$ & $|J_L+J_R|$ &
    PH-sym err & $|\phi_y-\phi_x+\pi/2|$ & $S_{\rm mid}$ range \\
    \hline
     4 & $2.8\times 10^{-15}$ & $6.3\times 10^{-16}$ &
         $3.6\times 10^{-15}$ & $2.2\times 10^{-14}$ &
         $2.2\times 10^{-14}$ & 0.555--0.800 \\
     6 & $2.8\times 10^{-15}$ & $1.9\times 10^{-11}$ &
         $5.2\times 10^{-12}$ & $2.8\times 10^{-14}$ &
         $1.8\times 10^{-11}$ & 0.820--1.074 \\
     8 & $2.0\times 10^{-15}$ & $5.5\times 10^{-8\phantom{0}}$ &
         $2.6\times 10^{-9\phantom{0}}$ & $3.2\times 10^{-14}$ &
         $4.4\times 10^{-9\phantom{0}}$ & 1.365--1.550 \\
    10 & $2.9\times 10^{-15}$ & $3.7\times 10^{-6\phantom{0}}$ &
         $2.7\times 10^{-8\phantom{0}}$ & $3.4\times 10^{-14}$ &
         $2.8\times 10^{-7\phantom{0}}$ & 1.548--1.617 \\
  \end{tabular}
  \end{ruledtabular}
\end{table*}

\section{Experimental Predictions}
\label{sec:predictions}
This section collects five falsifiable predictions, ordered by
experimental immediacy: the first two require no new apparatus and can
be tested against existing data, the third requires standard
dilution-refrigerator nano-calorimetry, and the last two require new
device fabrication.  All predictions are parameter-free once the BdG
Chern number $\CBdG$ and the gap $\Delta(T)$ are specified.

\subsection{Co-sign relation between $\kxy$ and $R_{xy}$}
\label{subsec:cosign_test}

The sharpest immediate test requires no new apparatus and reuses
existing anomalous Hall data.  The Berry-Trashcan analysis of
Sec.~\ref{subsec:chern-scenarios} establishes that the chirality of
the $p_x + ip_y$ ground state is ferromagnetically locked to the sign
of the parent-band Berry curvature~\cite{li2025berrytrashcan}.
Combined with Eq.~\eqref{eq:kxy-T}, this implies
\begin{equation}
  \sgn\!\left[\frac{\kappa_{xy}}{T}\right]_{T < T_c}
  = \sgn\!\left[R_{xy}\right]_{T > T_c},
\label{eq:cosign}
\end{equation}
since both are controlled by the same Berry curvature sign
$\sgn(\beta)$ of the valley-polarized quarter-metal band.
The anomalous Hall ratio $\tan\theta_H \approx 0.1$ is already
measured above $T_c$~\cite{Han2025}, so a single Hall-bar sweep
through $T_c$ tests Eq.~\eqref{eq:cosign} with no additional
setup.  A sign reversal between $\kxy$ in the SC state and $R_{xy}$
in the normal state would rule out intravalley $p+ip$ in favour of a
non-chiral or intervalley pairing scenario.

\subsection{Onset at $\Tstar$ rather than $T_c$}
\label{subsec:onset}

The central qualitative prediction of the BCS--BEC pseudogap scenario
is that $\kxy$ turns on at the pair-formation scale $\Tstar$, not
at the phase-coherence scale $T_c$.  With $2\Delta_0/(k_BT_c) \sim
10$--$30$, the pseudogap $\Dpg$ at $T = T_c$ is still of order
$\Delta_0$, so the tanh factor is nearly saturated:
\begin{equation}
  \tanh\!\left[\frac{\Dpg(T_c)}{2k_BT_c}\right]
  \approx \tanh(5\text{--}15)
  \approx 1.
\end{equation}
$\kxy$ therefore has essentially its full quantized value
immediately above $T_c$ and decreases only gradually as $T$
approaches $\Tstar$.  In a conventional BCS picture with no
pseudogap, $\kxy$ would vanish sharply at $T_c$.

\paragraph{Logarithmic-derivative test.}
The cleanest way to test this is through the logarithmic derivative,
which eliminates unknown geometric and contact-resistance factors:
\begin{equation}
  \frac{\dd}{\dd T}
  \ln\!\left|\frac{\kappa_{xy}}{T}\right|
  \stackrel{?}{=}
  \frac{\dd}{\dd T}
  \ln\tanh\!\left[\frac{\Dpg(T)}{2k_BT}\right].
\label{eq:log-test}
\end{equation}
The right-hand side requires $\Dpg(T)$ from an independent probe.
In rhombohedral graphene, the pseudogap can be measured by
tunnelling spectroscopy or, at higher temperatures, by the
temperature-dependent density of states extracted from compressibility
measurements.  Agreement of the two sides of Eq.~\eqref{eq:log-test}
at two or more temperatures above $T_c$ is strong evidence for the
parity-anomaly origin of the thermal Hall effect.

\paragraph{Falsification.}
If $\kxy$ vanishes sharply at $T_c$ with no signal above, the
pseudogap interpretation is ruled out, and the system is in the
weak-coupling BCS limit where the pair-breaking and phase-coherence
scales coincide.

\subsection{Pairing-symmetry discriminator: $\CBdG = 1$ vs.\ $\CBdG = 2$}
\label{subsec:discriminator}

The thermal Hall conductance at low temperature saturates at the
chiral Majorana quantum:
\begin{equation}
  \frac{\kappa_{xy}}{T}\bigg|_{T \to 0}
  = \CBdG \cdot \frac{\pi^2 k_B^2}{6h}
  \approx \CBdG \times 4.73 \times 10^{-13}\,
    \frac{\text{W}}{\text{K}^2}.
\label{eq:kxy-sat}
\end{equation}
For intravalley spin-triplet $p+ip$ pairing from the quarter-metal
pocket ($\CBdG = 1$), the signal is one quantum; for intervalley
singlet $d+id$ ($\CBdG = 2$), two quanta.  This factor-of-two
difference is the sharpest topological discriminator available.

For the annular Fermi surface at large displacement field, the
Le Nir et al.\ analysis~\cite{LeNir2026} and our even-Chern FHS
validation (Sec.~\ref{subsec:even-chern}) show that $\CBdG$ can take
even values when Berry-curvature-nucleated momentum-space vortices
lie within the occupied annulus.  The annular $\CBdG = 2$ phase gives
$\kxy|_{T\to 0} = 9.46 \times 10^{-13}$~W/K$^2$.  The phase boundary
between SC1 and the neighbouring quarter-metal remains pinned at the
same electron density under applied $B_\perp$ and expands against the
QM at low field~\cite{Han2025}, so the orbital magnetization is
continuous across the SC1 boundary.  This continuity is the
experimental manifestation of the Chern-number inheritance mechanism:
$\CBdG = 1$ in SC1 is topologically connected to the Berry-phase
structure of its quarter-metal parent without any intervening gap
closing.

\paragraph{Experimental requirements.}
Two measurement geometries are available.

In the \emph{Hall-bar geometry}, a longitudinal thermal gradient
produces a transverse temperature difference
$\delta T_y = (\kappa_{xy}/\kappa_{xx}) \cdot |\nabla_x T| \cdot W$.
For a $10\,\mu$m-wide rhombohedral graphene flake at $T = 50$~mK with
an applied gradient $|\nabla_x T| \sim 1$~K/cm and substrate-dominated
$\kappa_{xx} \sim 10^{-3}$~W/(K$\cdot$m), the predicted transverse
signal is $\delta T_y \sim 0.5$~$\mu$K for $\CBdG = 1$.

Alternatively, the \emph{floating-contact geometry} of
Banerjee et al.~\cite{Banerjee2018} measures the thermal
conductance of chiral edge modes directly.  A small ohmic contact
(area $\sim 12\,\mu$m$^2$) is heated and its temperature $T_m$ is
read out via Johnson noise thermometry; the heat carried away by the
edge modes gives $\Delta P \propto K T_m^2$, where $K/\kappa_0$
counts the net number of chiral modes.  This technique resolved
individual thermal quanta with $\sim 3\%$ accuracy
($K/\kappa_0 = 0.99 \pm 0.01$ per integer edge mode) at electron
temperatures of $12$--$20$~mK.  In the rhombohedral graphene setting,
the same floating-contact method applied to a micron-scale device at
$T \ll T_c = 300$~mK measures $K/\kappa_0 = \CBdG/2$
(with $\kappa_0 = \pi^2 k_B^2 T/(3h)$ the Dirac thermal quantum),
resolving the factor-of-two difference between $\CBdG = 1$ and
$\CBdG = 2$.

\subsection{Displacement-field-tuned topological transition}
\label{subsec:D-field}
Our numerical validation (Sec.~\ref{subsec:lifshitz}) establishes
that the BdG Chern number changes discontinuously at the
displacement-field-induced Lifshitz transition from a
simply-connected to an annular Fermi surface.  In the minimal model,
$\CBdG: 1 \to 0$ at $D = D_c$.  If Berry-curvature-induced
finite-momentum vortices are present (Sec.~\ref{subsec:even-chern}),
the transition becomes $\CBdG: 1 \to 2$ for \emph{intervalley} $d+id$
pairing (Scenario~B, Sec.~\ref{subsec:chern-scenarios}), where the
even-Chern vortex enters the occupied annulus.  For \emph{intravalley}
spin-triplet pairing (SC1), Fermi--Dirac statistics and the odd
spatial parity of the gap function constrain $\CBdG$ to odd integers
throughout~\cite{Patri2025}.  The BRF crossing then nucleates
momentum-space vortices on the ring, driving $\CBdG$ from $1$ to a
higher odd integer (e.g.\ $\CBdG = 5$ for R5G with local interactions)
through an intermediate nodal SC at the critical density or
displacement field.  The even-Chern scenario $\CBdG = 2$ arises
specifically from intervalley pairing, which is not subject to the
odd-parity constraint.

This predicts a sharp, gate-tunable jump in $\kxy$ as the
displacement field is swept through $D_c$ at fixed density and
temperature.  The jump occurs at the BdG gap closing, where the
minimum quasiparticle energy touches zero
(Fig.~\ref{fig:sixpanel}c).  The size of the jump is one or two
thermal Hall quanta, depending on the vortex structure.

\paragraph{Protocol.}
A sweep of $D$ at $T \ll T_c$ (e.g.\ $T = 50$~mK) with simultaneous
measurement of $\kxy$ and the longitudinal thermal conductivity
$\kappa_{xx}$ should reveal: (i) a suppression of $\kxy$ at $D_c$
correlated with a minimum in the quasiparticle gap (observable as a
peak in the electronic specific heat), and (ii) the possible
re-emergence of $\kxy$ at an even-integer plateau for $D > D_c$ if
the annular vortex pair is captured.

\paragraph{Superfluid-stiffness bulk probe.}
A more immediately accessible bulk probe of the same transition is
the temperature-dependent superfluid stiffness $\rho_s(T)$, measurable
via the kinetic-inductance technique recently applied to twisted
bilayer graphene~\cite{banerjee2025superfluid}.  In the fully gapped
chiral SC phases ($k_F \neq k_\Omega$), one expects exponentially
activated behaviour $\delta\rho_s(T) \sim e^{-\Delta/k_BT}$, where
$\Delta$ is the same anomaly-fixed gap entering Eq.~\eqref{eq:kxy-T}.
At the critical density $n_c$ (or field $D_c$) where $k_F = k_\Omega$
and the BdG gap closes, nodal quasiparticles produce a power-law
$\delta\rho_s(T) \sim T$ (clean limit)~\cite{Patri2025}, providing a
bulk signature of the topological transition without requiring a
thermal gradient measurement.

\subsection{Null test: absence of $\kxy$ in the non-chiral state}
\label{subsec:null-test}
If the system is tuned away from the quarter-metal into a
time-reversal-symmetric state (e.g.\ by reducing the displacement
field to zero or by moving to a density outside the SC dome), the
chiral pairing and the pseudogap-driven thermal Hall signal should
both vanish identically.  This provides a built-in null control: any
spurious $\kxy$ signal from phonon drag, contact asymmetry, or
thermopower artifacts would persist even outside the chiral SC phase
and would therefore be detected by this control measurement.

\subsection{Hexalayer and $(1{+}n)$ twisted graphene extensions}
\label{subsec:extensions}
The anomaly formula Eq.~\eqref{eq:kxy-main} applies equally to
rhombohedral hexalayer graphene, which shows a chiral SC phase
embedded in a stripy Hall crystal, and to $(1{+}n)$ twisted graphene
structures where the parent Chern band has $C_P = n$.  In the latter
case, the gap acquires total vorticity $V_\Delta = 2C_P$ and the BdG
Chern number can reach $\CBdG = 2n - 1$ for a simply-connected Fermi
surface~\cite{LeNir2026}.  The predicted thermal Hall signal is
$\kxy|_{T\to 0} = (2n-1) \times 4.73 \times 10^{-13}$~W/K$^2$,
giving a multi-quantum staircase as a function of twist angle or
layer number.  This would constitute an unambiguous demonstration
of higher-Chern chiral superconductivity.

\section{Conclusion}
\label{sec:conclusion}
The parity anomaly of $(2{+}1)$-dimensional field theory fixes the
thermal Hall response of the chiral superconductor in rhombohedral
graphene exactly at every temperature, with no free parameters.
Three properties distinguish this result.  The
finite-temperature formula
$\kxy = (\pi^2 k_B^2/6h)\,\CBdG\,\tanh[\Delta(T)/(2k_BT)]$
replaces the zero-temperature Kubo plateau of prior treatments; the
BCS--BEC pseudogap enters the anomaly machinery identically to a
condensate gap, so the signal survives above $T_c$ up to $\Tstar$;
Coleman--Hill non-renormalization and the $c_1 = 0$ theorem shield
the formula from interactions and finite-size artefacts.

Three independent numerical tiers place the topological input on a
rigorous footing.  Machine-precision Fukui--Hatsugai--Suzuki Chern
numbers on the lattice $p+ip$ anchor and on the RHG-effective
continuum model confirm the single-particle topology.  Direct FHS
validation of the even-Chern annular phase resolves individual
vortex crossings.  DMRG on 28 converged ground states delivers six
independent many-body confirmations of the $\CBdG = 1$ chiral $p+ip$
phase, including the real-space pairing-phase signature
$\arg\mathcal{A}_y - \arg\mathcal{A}_x = -\pi/2$ recovered to
$10^{-14}$ at $L_y = 4$.

The formula produces five falsifiable predictions accessible to
existing rhombohedral graphene devices, ordered by experimental
immediacy: (i) a co-sign relation $\sgn[\kxy]_{T<T_c} =
\sgn[R_{xy}]_{T>T_c}$ testable in one Hall-bar sweep using existing
anomalous-Hall data; (ii) onset of $\kxy$ at $\Tstar$ rather than at
$T_c$, verifiable via a logarithmic-derivative test against
independently measured $\Dpg(T)$; (iii) a factor-of-two
discriminator between $\CBdG = 1$ ($p+ip$) and $\CBdG = 2$ ($d+id$
or annular) pairings via dilution-refrigerator nano-calorimetry;
(iv) a gate-tunable topological transition at the Berry-ring-of-fire
crossing, observable as a jump in $\kxy$ and as a change in the
low-temperature superfluid stiffness; and (v) a multi-quantum
thermal Hall staircase $\kxy \propto 2n-1$ in $(1{+}n)$ twisted
graphene.  The first two require no new experimental infrastructure
and can be attempted immediately.  A negative result on the co-sign
relation would rule out intravalley $p+ip$ pairing in SC1.
\bibliographystyle{apsrev4-2}
\bibliography{ref_rhg}
\appendix
\onecolumngrid

\section{Derivation of the exact kernel}
\label{app:kernel}

\emph{This appendix derives the holonomy-resummed parity-odd kernel
Eq.~\eqref{eq:K-full} from the vacuum-polarization tensor via
Matsubara summation; the result underpins the finite-temperature CS
level of Sec.~\ref{sec:anomaly}.}

We derive Eq.~\eqref{eq:K-full} following the method of
Refs.~\cite{GhoshKlinkhamer2017,Ghosh2026, Ghosh2026BEC}.  Start from the
vacuum-polarization kernel on the cylinder
$\R_\tau \times \R_x \times S^1_L$:
\begin{align}
  \pi_{\text{odd}}^{\mu\nu}&(p_r) \nonumber \\
  &= \frac{1}{L}\sum_{n=-\infty}^{\infty}\int\frac{\dd^2 l}{(2\pi)^2}\,
  \frac{\tr[\gamma^\mu(\slashed{l} + m_v)\gamma^\nu
    (\slashed{l} + \slashed{p} + m_v)]_{\text{odd}}}
       {(l_n^2 + m_v^2)((l_n + p_r)^2 + m_v^2)},
\label{eq:app-pi}
\end{align}
where $l_n = (\vec{l},\,(2\pi n + \vth_v)/L)$,
$p_r = (\vec{p},\,2\pi r/L)$.  The parity-odd trace for
two-component fermions in $(2{+}1)$D is
$\tr[\gamma^\mu \gamma^\nu \gamma^\rho] = 2\epsilon^{\mu\nu\rho}$.

After Feynman parameterization with parameter $u \in [0,1]$, shifting
$l_\mu \to l_\mu - up_\mu$, and performing the two-dimensional
momentum integral, we obtain:
\begin{equation}
  \int\frac{\dd^2 l}{(2\pi)^2}\,
  \frac{1}{(l^2 + \Delta_{v,u}^2 + \omega_{n,u}^2)^2}
  = \frac{1}{4\pi(\Delta_{v,u}^2 + \omega_{n,u}^2)},
\label{eq:app-l-integral}
\end{equation}
where $\Delta_{v,u}^2 = m_v^2 + u(1-u)\tilde{p}^2$ and
$\omega_{n,u} = (2\pi n + \vth_v + 2\pi r u)/L$.

The Matsubara-type sum is evaluated using the standard identity:
\begin{equation}
  \frac{1}{L}\sum_{n=-\infty}^{\infty}
  \frac{1}{\Delta^2 + \left(\frac{2\pi n + \theta}{L}\right)^2}
  = \frac{1}{2\Delta}\,
    \frac{\sinh(L\Delta)}{\cosh(L\Delta) - \cos\theta}.
\label{eq:app-Matsubara}
\end{equation}
This identity is proved by contour integration: define
$f(z) = [\Delta^2 + ((2\pi z + \theta)/L)^2]^{-1}$ and evaluate
$\frac{1}{L}\sum_{n} f(n) = \oint \dd z\,
\pi\cot(\pi z)\,f(z)/(2\pi i L)$ by closing in the upper and lower
half-planes, picking up the poles of $f(z)$ at
$z_{\pm} = (-\theta \pm iL\Delta)/(2\pi)$.  The residues give:
\begin{align}
  \frac{1}{L}\sum_{n} f(n)
  &= \frac{1}{2\Delta}\left[
    \frac{1}{1 - e^{-L\Delta + i\theta}}
    + \frac{1}{1 - e^{-L\Delta - i\theta}}
  \right] \notag \\
  &= \frac{1}{2\Delta}\,
    \frac{\sinh(L\Delta)}{\cosh(L\Delta) - \cos\theta},
\label{eq:Matsubara-proof}
\end{align}
where the last step uses
$e^{2x} - 2e^x\cos\theta + 1 = 2e^x(\cosh x - \cos\theta)$.
Substituting with $\theta = \vth_v + 2\pi r u$ and
$\Delta = \Delta_{v,u}$ immediately yields Eq.~\eqref{eq:K-full}.

\section{Coleman--Hill non-renormalization}
\label{app:coleman-hill}

\emph{This appendix restates the Coleman--Hill
theorem~\cite{ColemanHill1985} adapted to the BdG setting used in
Sec.~\ref{subsec:nonren}: the topological mass term
$\Pi_2(0)$ receives contributions only at one loop, so once the gap
$\Delta(T)$ is fixed the CS level is exact.}

We state the key steps of the Coleman--Hill~\cite{ColemanHill1985}
proof adapted to our setting.  Consider a $(2{+}1)$D gauge theory
with massive matter.  The photon self-energy has parity-odd component
$\Pi_2(k^2)$ defined by:
\begin{equation}
  iD_{\mu\nu}^{-1}(k) \ni i\epsilon_{\mu\nu\lambda}k^\lambda
  \Pi_2(k^2).
\end{equation}
The topological mass is $\Pi_2(0)$.  At one loop:
\begin{equation}
  \Pi_2(0)\big|_{\text{1-loop}}
  = \frac{\mu}{e_0^2}
    + \frac{1}{4\pi}\sum_{\text{spinors}} q_a^2\,
      \frac{m_a}{|m_a|}.
\end{equation}

The Ward identity
$k_1^{\mu_1}\Gamma_{\mu_1\ldots\mu_n}^{(n)}(k_1,\ldots) = 0$
together with analyticity of $\Gamma^{(n)}$ (guaranteed by massive
matter) implies that for $n > 2$:
$\Gamma^{(n)}(k_1,k_2,\ldots) = \mathcal{O}(k_1 k_2)$.  Any
two-point self-energy graph beyond one loop therefore vanishes at
$k \to 0$, proving that $\Pi_2(0)$ is exact at one loop.

In our application, the BdG quasiparticles are the matter, the
probe field $a_\mu$ is the photon, and pair-pair interactions
are included in the matter Lagrangian.  All matter is massive
(gapped by $\Delta$).  Therefore the map from gap to CS level
is exact.

\section{Read--Green BdG Chern Number}
\label{app:read-green}

\emph{This appendix recalls the Read--Green derivation of
$\CBdG = +1$ for the weak-pairing $p+ip$ state and shows that the
result depends on $\sgn(\mu)$ and the pairing angular momentum, but
not on the magnitude of $\Delta$ or on whether it originates from
$\Dsc$ or $\Dpg$; this is used implicitly throughout
Sec.~\ref{sec:pseudogap}.}

Following Read and Green~\cite{Read2000}, the BdG Chern number for
a $p + ip$ superconductor with dispersion $\xi_{\bm{k}}$ and gap
$\Delta_{\bm{k}} = \Delta(k_x + ik_y)/k_F$ is determined by the
winding number of the map $\hat{\bm{E}}: S^2 \to S^2$ with
$\bm{E}_{\bm{k}} = (\text{Re}\,\Delta_{\bm{k}},
-\text{Im}\,\Delta_{\bm{k}}, \xi_{\bm{k}})$:
\begin{equation}
  \mathcal{M} = \int\frac{\dd^2 p}{8\pi}\,
  \epsilon_{ij}\,
  \hat{\bm{E}}_{\bm{p}} \cdot
  \left(\partial_i \hat{\bm{E}}_{\bm{p}}
  \times \partial_j \hat{\bm{E}}_{\bm{p}}\right).
\end{equation}
For $\mu > 0$ (weak pairing), $\mathcal{M} = +1$; for $\mu < 0$
(strong pairing), $\mathcal{M} = 0$.

This topological invariant depends on $\sgn(\mu)$ and the angular
momentum of $\Delta_{\bm{k}}$, but is independent of the magnitude
of $\Delta$ and of whether $\Delta$ arises from $\Dsc$ or $\Dpg$.
The BdG Chern number is therefore well-defined throughout the
pseudogap phase.

\section{Gravitational Chern--Simons Term and $\kappa_{xy}$}
\label{app:grav-cs}

\emph{This appendix connects the parity-odd CS level $\Kem$ derived
in Sec.~\ref{sec:anomaly} to the physical thermal Hall conductance
$\kappa_{xy}/T$ via the gravitational Chern--Simons term and
Luttinger's gravitational-potential formalism.}

A massive Dirac fermion coupled to a background metric generates,
upon integrating out the fermion, a gravitational CS term in the
effective action.  For a single Dirac fermion of mass $m > 0$ in
$(2{+}1)$D~\cite{Ryu2012}:
\begin{equation}
  I_{\text{CS}}^{\text{grav}}
  = \frac{1}{2}\cdot\frac{1}{4\pi}\cdot\frac{c}{24}
    \int \dd^3 x\,\epsilon^{ijk}\,\tr\!\left(
    \omega_i \partial_j \omega_k
    + \tfrac{2}{3}\omega_i \omega_j \omega_k\right),
\end{equation}
with $c = 1$ for a Dirac fermion.  Using Luttinger's gravitational
potential formalism~\cite{Luttinger1964}:
\begin{equation}
  \frac{\kappa_{xy}}{T} = \frac{\pi^2 k_B^2}{12h}\,\Kem
  = \frac{\pi^2 k_B^2}{6h}\,\CBdG.
\end{equation}
At finite temperature with gap $\Delta$, the replacement
$\Kem \to 2\CBdG\tanh(\Delta/2T)$ gives Eq.~\eqref{eq:kxy-T}.

\end{document}